\documentclass[twocolumn,twocolappendix]{aastex63}

\shortauthors{Duan \& Guo 2024}

\begin{document}
\shorttitle{Cold Filaments Formed in Hot Wake Flows of AGN Bubbles}
\title{Cold Filaments Formed in Hot Wake Flows Uplifted by Active Galactic Nucleus Bubbles in Galaxy Clusters}


\author[0000-0002-6921-1899]{Xiaodong Duan}
\affiliation{School of Physics, Henan Normal University, Xinxiang 453007, China; duanxiaodong@htu.edu.cn}

\author[0000-0003-1474-8899]{Fulai Guo}
\affiliation{Key Laboratory for Research in Galaxies and Cosmology, Shanghai Astronomical Observatory, Chinese Academy of Sciences\\
80 Nandan Road, Shanghai 200030, China; fulai@shao.ac.cn}
\affiliation{School of Astronomy and Space Science, University of Chinese Academy of Sciences,
19A Yuquan Road, Beijing 100049, China}
\affiliation{Tianfu Cosmic Ray Research Center, 610000 Chengdu, Sichuan, China}

\begin{abstract}

Multi-wavelength observations indicate that the intracluster medium in some galaxy clusters contains cold filaments, while their formation mechanism remains debated. Using hydrodynamic simulations, we show that cold filaments could naturally condense out of hot gaseous wake flows uplifted by the jet-inflated active galactic nucleus (AGN) bubbles. Consistent with observations, the simulated filaments extend to tens of kiloparsecs from the cluster center, with a representative mass of $\rm 10^{8}- 10^{9}\ M_{\odot}$ for a typical AGN outburst energy of $10^{60}~ \rm erg$. They show smooth velocity gradients, stretching typically from inner inflows to outer outflows with velocity dispersions of several hundred kilometers per second. The properties of cold filaments are affected substantially by jet properties. Compared to kinetic-energy-dominated jets,  thermal-energy-dominated jets are easier to produce long cold filaments with large masses as observed. AGN jets with an early turn-on time, a low jet base, or a very high power tend to overheat the cluster center, and produce short cold filaments that take a relatively long time to condense out. 

\end{abstract}


\section{Introduction}
\label{section1}

Multi-wavelength observations show that the intracluster medium (ICM) in some galaxy clusters contains cold filaments with $\rm H_{\alpha}$ and $\rm CO$ emissions, but how these cold filaments arise in the ICM remains a puzzle (\citealt{conselice01}; \citealt{salom06}; \citealt{tamhane23}; \citealt{oosterloo23}). These filaments often extend to several tens kpc from the cluster center (\citealt{mcdonald09}; \citealt{calzadilla22}), and have typical masses of $\rm 10^{8} - 10^{10}\ M_{\sun}$,  width $\rm < 1\ kpc$, and smooth velocity gradients (\citealt{hatch06}; \citealt{salom08}; \citealt{mcdonald09}; \citealt{mcdonald12}; \citealt{hamer16}; \citealt{fabian16}; \citealt{olivares19}; \citealt{russell19}; \citealt{olivares19}). Furthermore, observations show that some elongated cold filaments lie between active galactic nucleus (AGN) bubbles and the cluster center (\citealt{fabian03}; \citealt{salom08}; \citealt{russell17}), a phenomenon sometimes named as the `alignment effect' in radio galaxies (\citealt{chambers87}; \citealt{mccarthy87}; \citealt{best96}; \citealt{saikia22}). This leads to the conjecture that these cold filaments may be uplifted by AGN bubbles inflated by AGN jets (\citealt{fabian03}; \citealt{mcdonald10}; \citealt{churazov13}; \citealt{mcnamara16}; \citealt{calzadilla22}). Though AGN feedback is believed to be an important heating source for solving the cooling flow problem in the ICM (see \citealt{fabian12}; \citealt{mcnamara12}; \citealt{donahue22}; \citealt{bourne23} for reviews), the physical relation between AGN bubbles and cold filaments has not been firmly established, and should be investigated in combination with the complex gas dynamic processes of AGN feedback (\citealt{guo18}). 

Theoretical analyses and numerical simulations show that AGN bubbles could drag the hot ICM away from the cluster center. \citet{pope10} propose the Darwin drift model in fluid mechanics to explain the filaments behind buoyant AGN bubbles (\citealt{darwin53}; \citealt{dabiri06}). Hydrodynamic simulations confirm the formation of extended hot wake flows behind AGN bubbles by evolving existing or instantaneously generated off-center low-density bubbles (\citealt{churazov01}; \citealt{saxton01}; \citealt{bruggen03}; \citealt{reynolds05}; \citealt{zhang22}). In \citet{revaz08} and \citet{brighenti15}, a fraction of the hot gas embedded in the wake flows cools down to $10^{4}$ K, forming cold filaments. These studies focus on the buoyant evolution of off-center AGN bubbles, without directly simulating the process of jet evolution. In more realistic cases, the formation of AGN bubbles during the jet evolution and the associated shock heating processes should also be simulated to study this mechanism, as shocks heat the ICM, potentially preventing the formation of cold filaments. The jet simulations of \citet{duan18} show that hot wake flows (referred as "trailing outflows" therein) indeed form behind the jet-inflated bubbles, roughly consistent with the Darwin drift model predictions. While potentially explaining hot metal-rich outflows preferentially aligned with the large-scale axes of X-ray cavities in observations (\citealt{simionescu09}; \citealt{kirp11}; \citealt{kirp15}), hot wake flows do not cool down to form cold filaments in \citet{duan18}.  

The correlation between cold gas condensation and AGN feedback in the ICM has been investigated in many studies. Some authors find that local thermal instability leads to cold gas condensation when the environmental gas is significantly disturbed by AGN outflows (\citealt{gaspari12}; \citealt{li14}; \citealt{valentini15}; \citealt{beckmann19}), while others directly initiate multi-phase AGN outflows containing relatively cold gas at the cluster center (\citealt{qiu20}). In these studies, AGN feedback is often modelled with multi-phase outflows ab initio, while cold gas condensation is usually clumpy and spatially widespread but not concentrated along the jet axis as those observed thin filaments. Therefore, cold gas condensation in these simulations may not be the main formation mechanism of those observed cold filaments stretched linearly between AGN bubbles and the cluster center (\citealt{conselice01}; \citealt{fabian03}; \citealt{hatch06}; \citealt{salom08}; \citealt{mcdonald09}; \citealt{fabian16}; \citealt{olivares19}).

In this work, we focus on the formation of linear cold filaments condensed out of hot wake flows uplifted by jet-inflated AGN bubbles, following our previous work on hot metal-rich wake flows of AGN bubbles in the ICM (\citealt{guo18}; \citealt{duan18}). In \citet{duan18}, hot gas in wake flows does not cool down to form cold filaments, probably due to the relatively low spatial resolution (hundreds pc) adopted. Here we greatly increase spatial resolution with the adaptive mesh refinement (AMR) technique. To simulate AGN feedback, it is important to properly choose the set-up of jet injection and particularly jet parameters. While the matter and energy contents of AGN jets are still unclear observationally (\citealt{boehringer93}; \citealt{fan18}; \citealt{blandford19}; \citealt{hardcastle20}), simulations of AGN jet outbursts with various jet parameters show intriguing diversities in bubble morphology and heating processes (\citealt{guo15}; \citealt{sukungyi21}; \citealt{weinberger23}; \citealt{husko23a}; \citealt{husko23b}). A kinetic-energy dominated jet preferentially inflates radially-elongated bubbles far from the cluster center and deposits energy there, while a thermal-energy dominated jet tends to inflate transverse-wide bubbles and deposit energy closer to the cluster center (\citealt{duan20}; \citealt{guo20}). Here we also investigate how the energy content of AGN jets influences the formation and properties of cold filaments.

The remainder of the paper is organized as follows. In Section \ref{section2}, we analytically estimate the mass of wake flows uplifted by AGN bubbles in the ICM. In Section \ref{section3}, we present a series of simulations with varying jet parameters to study the formation of cold filaments in hot wake flows uplifted by AGN bubbles in the ICM. Finally, in Section \ref{section4}, we present a summary of our results with some brief discussions.


\section{Analytical Estimations}
\label{section2}

In this section, we estimate the typical mass of hot wake flows uplifted by AGN bubbles via the Darwin drift model (\citealt{darwin53}; \citealt{dabiri06}), which serves as an upper limit of the mass of cold filaments forming in wake flows behind AGN bubbles. As an object moves through an ambient fluid, the Darwin drift refers to a net displacement of some fluid behind the object along its moving direction (\citealt{darwin53}). Though bubbles inflated by AGN jets are not ideal spherical objects and the ambient ICM is not potential flow as considered in the original Darwin drift model (\citealt{darwin53}), we will apply this model to derive a rough estimate of the wake mass and show that the result is well consistent with previous simulations (\citealt{duan18}) and observations (\citealt{kirp15}).   

In the Darwin drift model, the drift volume $V_{\rm drift}$ can be estimated as (\citealt{darwin53}; \citealt{dabiri06}):
\begin{eqnarray} 
V_{\rm drift}=k_{\rm d}V_{\rm body}\label{eqdrift}
\end{eqnarray}
where $V_{\rm body}$ is the volume of the moving body (e.g., AGN bubble), and $k_d$ is a numerical constant with a value of $0.5$ for the case of a moving spherical solid object (\citealt{darwin53}) or $0.72$ for the case of a moving vortex bubble (\citealt{dabiri06}).

We may estimate the jet outburst energy $E_{\rm jet}$ from X-ray observations of galaxy clusters according to \citep{duan20}
\begin{eqnarray} 
E_{\rm jet} \approx \frac{p_{\rm bub}V_{\rm bub}}{(1-\eta_{\rm cp})(\gamma_{\rm bub}-1)}\label{eqEjet} ,
\end{eqnarray}
where $p_{\rm bub}$ and $V_{\rm bub}$ are the observed pressure and volume of AGN bubbles or X-ray cavities, respectively, and $\gamma_{\rm bub}$ is the effective adiabatic index of the plasma in AGN bubbles. $\eta_{\rm cp}$ is  the energy coupling efficiency, defined as the fraction of AGN jet energy transferred to the ICM, typically around $0.7-0.9$ for both powerful and weak jet outbursts, as shown in \citet{duan20} with hydrodynamic jet simulations. For relativistic plasma, taking $\gamma_{\rm bub}=4/3$, the outburst energy can be estimated as $E_{\rm jet} \approx 10 - 30\ p_{\rm bub}V_{\rm bub}$ for $\eta_{\rm cp} = 0.7 - 0.9$. If taking $\gamma_{\rm bub}=5/3$, $E_{\rm jet} \approx 5 - 15\ p_{\rm bub}V_{\rm bub}$ for $\eta_{\rm cp} = 0.7 - 0.9$. 

For a rough estimation, we assume that the wake flow (drift) behind an AGN bubble has a uniform density $\rho_{\rm wake}$, pressure $p_{\rm wake}$, and a total mass of $M_{\rm wake}$. Equation (\ref{eqdrift}) may be rewritten as 
\begin{eqnarray}  
V_{\rm bub}=\frac{M_{\rm wake}}{k_{\rm d} \rho_{\rm wake}} .
\label{eqVbub}
\end{eqnarray}
If we suppose the ratio of the pressure of the bubble to the wake flow is $\alpha_{\rm p}=p_{\rm bub}/p_{\rm wake}$, combining equation (\ref{eqVbub}) and equation (\ref{eqEjet}) and taking $p_{\rm wake} = \rho_{\rm wake} k_{\rm B } T_{\rm wake} / (\mu m_{\rm p})$, we have
\begin{eqnarray} 
M_{\rm wake} = \frac{k_{\rm d}(1-\eta_{\rm cp})(\gamma_{\rm bub}-1)\mu m_{\rm p}E_{\rm jet}}{\alpha_{\rm p} k_{\rm B}T_{\rm wake}} ,
\end{eqnarray}
where $k_{\rm B}$ is the Boltzmann constant, $m_{p}$ is the proton mass, $T_{\rm wake}$ is the typical gas temperature in the wake flow, $\mu$ is the mean molecular weight per particle, and we take $\mu=0.61$. Thus, we have
\begin{eqnarray} 
M_{\rm wake} \approx 2.67 \times 10^{10}\frac{E_{\rm jet}}{10^{60} \rm erg} \left( \frac{T_{\rm wake}}{10^7 \rm K}\right)^{-1}\rm M_{\odot}.
\label{eqmwake}
\end{eqnarray}
Here we have taken $k_{\rm d}=0.72$, $\gamma_{\rm bub}=4/3$, $\eta_{\rm cp} = 0.7$, $\alpha_{\rm p}=1$, $T_{\rm wake}$ with a typical value of $10^7$K in the cluster inner regions, and the energy of the AGN jet outburst with a typical value of $10^{60} $erg. This estimation of the mass of the wake flow uplifted by an AGN bubble is consistent with X-ray observations of metal-rich outflows (\citealt{kirp15}) and our previous simulation (\citealt{duan18}).

In Equation (\ref{eqmwake}), we give an estimate of the mass of wake flows but not that of cold filaments. In our model, cold filaments form via radiative cooling in hot wake flows, and thus, a cold filament uplifted by an AGN bubble should have a mass lower than our estimated value here. We investigate this problem with hydrodynamic simulations in the following section.  


\section{Hydrodynamic Simulations}
\label{section3}

In this section, we present a suite of hydrodynamic simulations to explore the formation of cold filaments in wake flows uplifted by AGN bubbles in the ICM. We focus on a thermal-energy-dominated jet run and a kinetic-energy-dominated jet run, comparing the differences in filament formation processes. We further investigate the impact of other model parameters on our results in Section \ref{section3.5}.

\subsection{Numerical Setup}
\label{section3.1}

\begin{deluxetable}{lccccc}
\label{tab1}
 \centering
\tablenum{1}
\tablecaption{List of Our Jet Simulations}
\tablehead{
\colhead{Run} & \colhead{$\rm f_{ th}$} &  \colhead{$\rm t_{ jeton}/Myr$} & \colhead{$\rm t_{ jet}/Myr $}  & \colhead{ $\rm z_{ jet} / kpc $}
}
\startdata
Jfth9            &     0.9        &   250     &   50	&     5             \\
Jfth1            &      0.1        &   ---      &    ---         &    ---              \\
Jtjeton200   &      ---        &   200     &    ---         &    ---              \\
Jtjeton300   &      ---        &   300     &    ---         &    ---              \\
Jtjet5           &     ---         &   ---       &   5   &     ---            \\
Jzjet1          &      ---        &    ---      &   ---         &    1           \\
\enddata
\tablecomments{The relevant jet parameters in our simulations include the thermal energy fraction at the jet base $\rm f_{th}$, the jet turn-on time $\rm t_{ jeton}$, the jet duration $\rm t_{jet}$ and the height of the jet base $\rm z_{jet}$. The mark '---' means that the corresponding parameter has the same value as in the fiducial run Jfth9. }   
\end{deluxetable}

\begin{figure*}
\gridline{
\includegraphics[height=0.316\textheight]{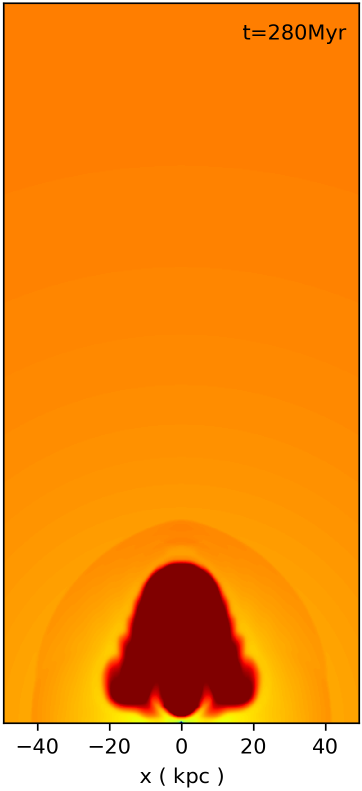}
\includegraphics[height=0.316\textheight]{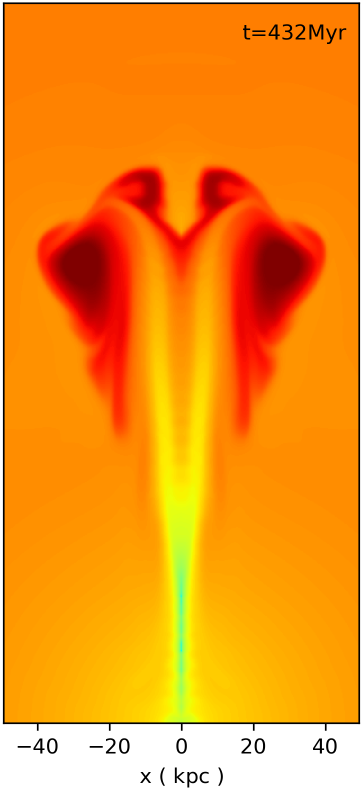}
\includegraphics[height=0.316\textheight]{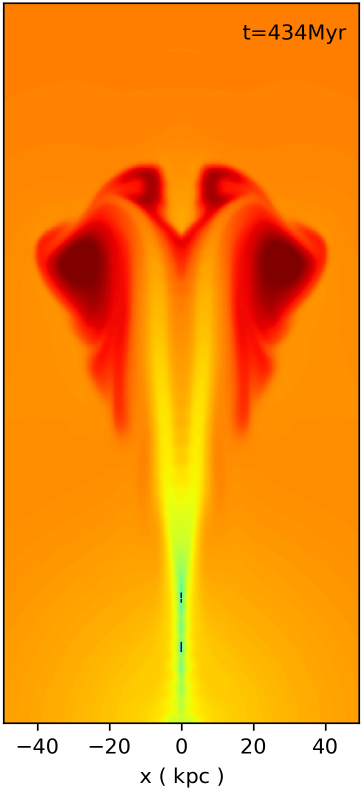}
\includegraphics[height=0.316\textheight]{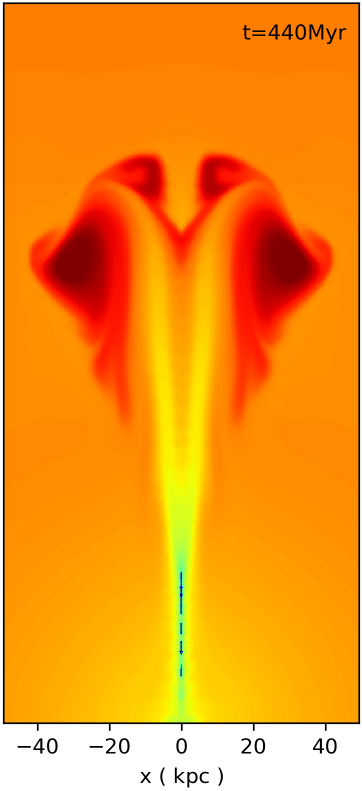}
\includegraphics[height=0.319\textheight]{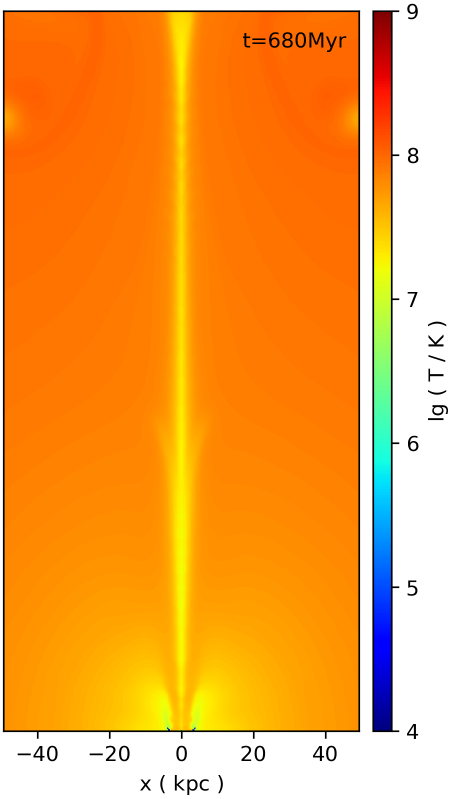} 
}
\gridline{
\includegraphics[height=0.340\textheight]{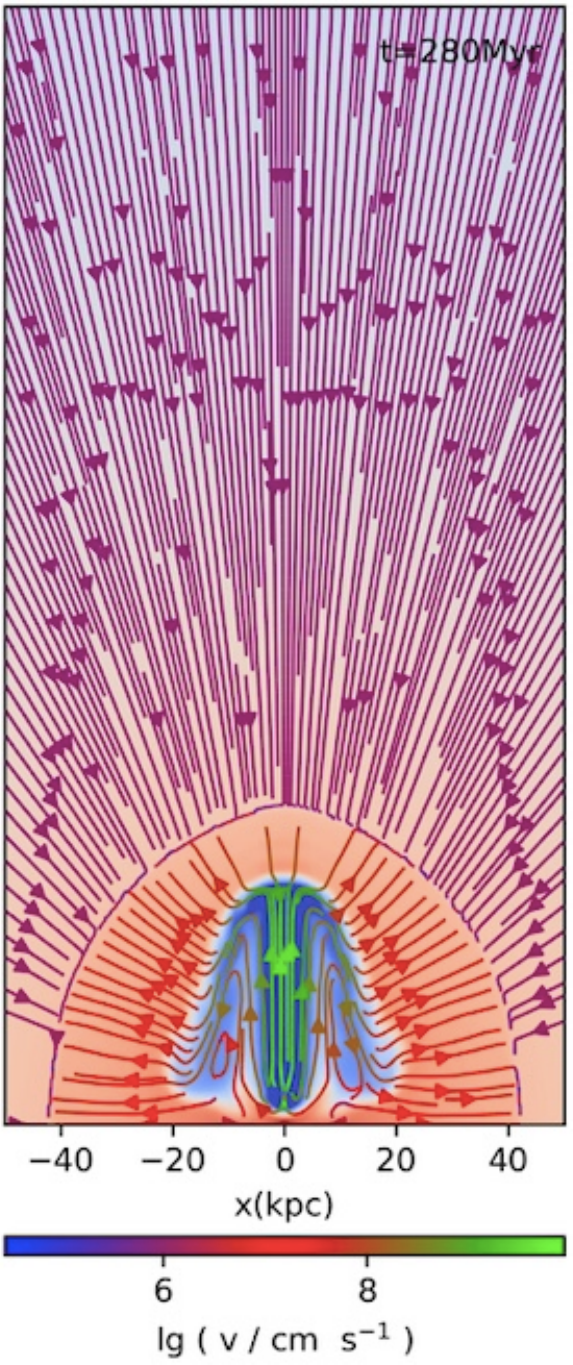}
\includegraphics[height=0.340\textheight]{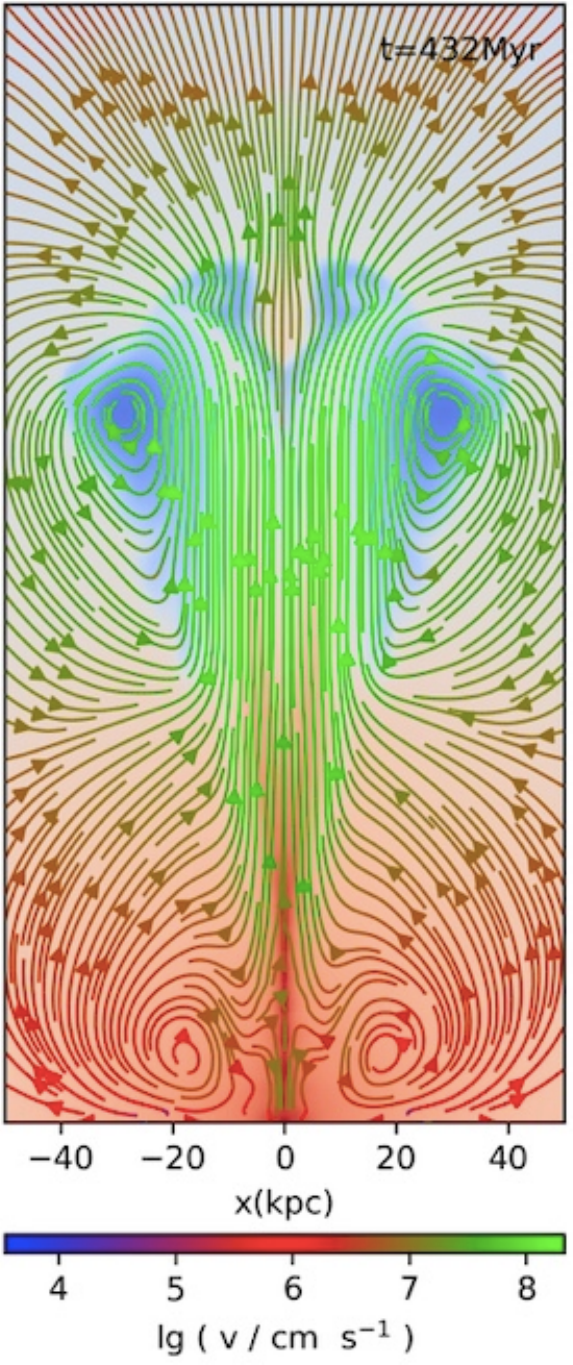}
\includegraphics[height=0.340\textheight]{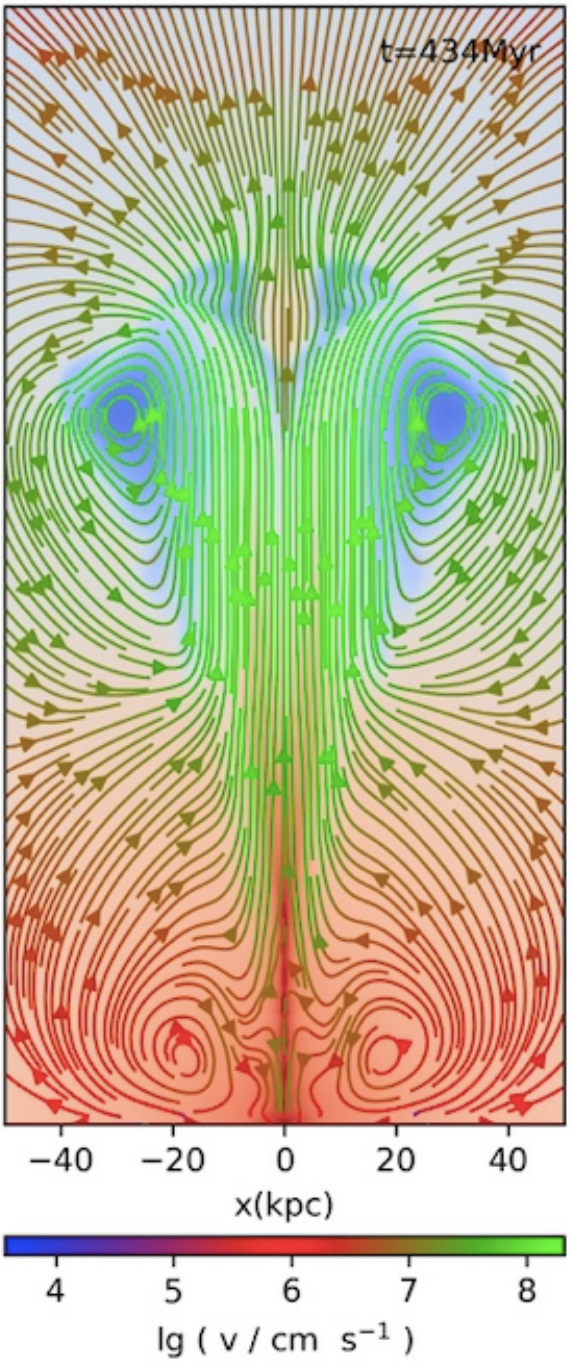}
\includegraphics[height=0.340\textheight]{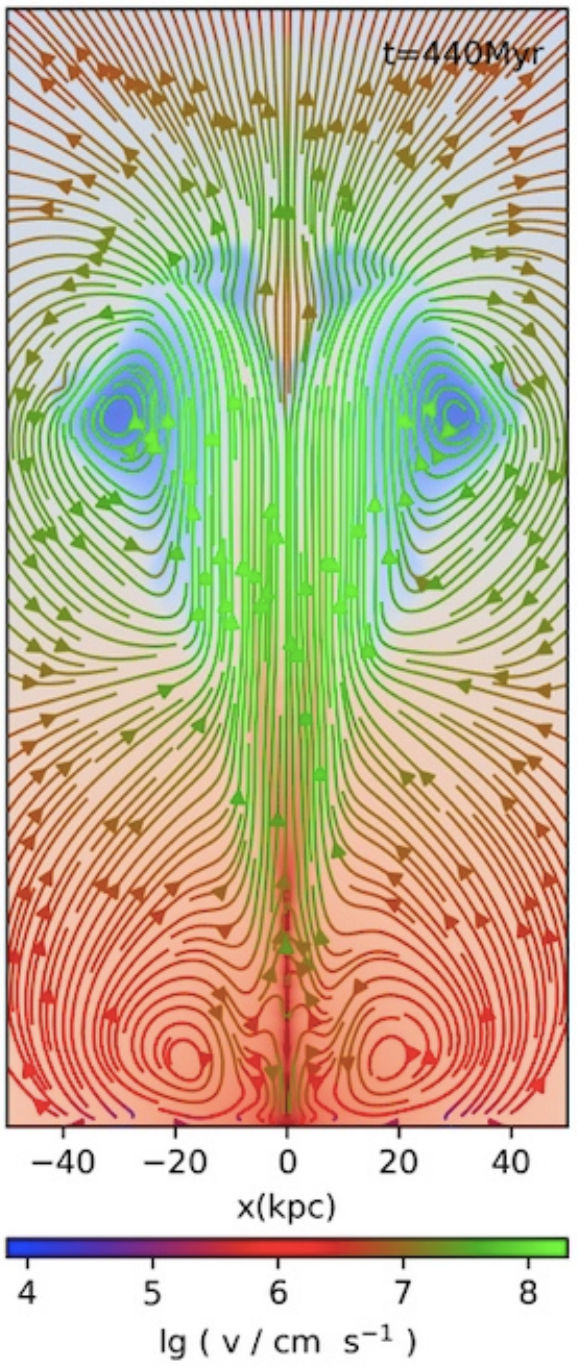}
\includegraphics[height=0.343\textheight]{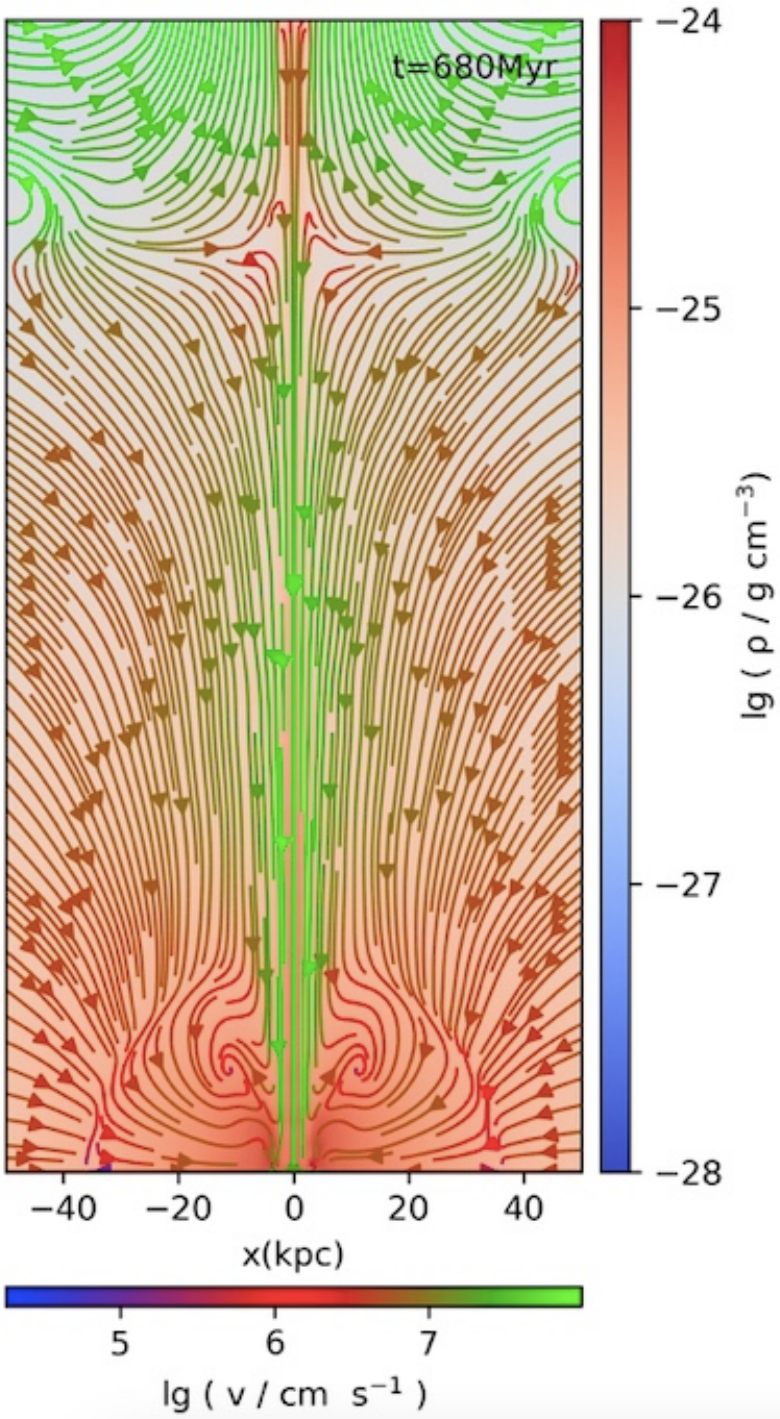}
}
\caption{Top panels: temporal evolution of the gas temperature distribution in our fiducial run Jfth9 with a thermal-energy-dominated jet (see Table \ref{tab1} for the adopted jet parameters); Bottom panels: gas velocity streamlines overlapping with the gas density distribution in run Jfth9.}
\label{figtv-fth9}
\end{figure*}

\begin{figure*}[!ht] 
\centering
\gridline{
\includegraphics[height=0.316\textheight]{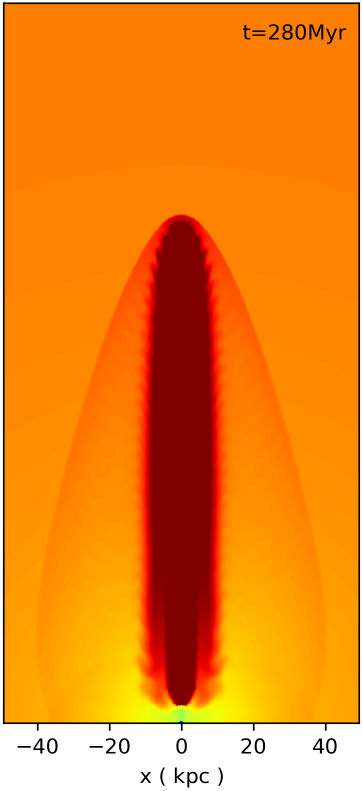}
\includegraphics[height=0.316\textheight]{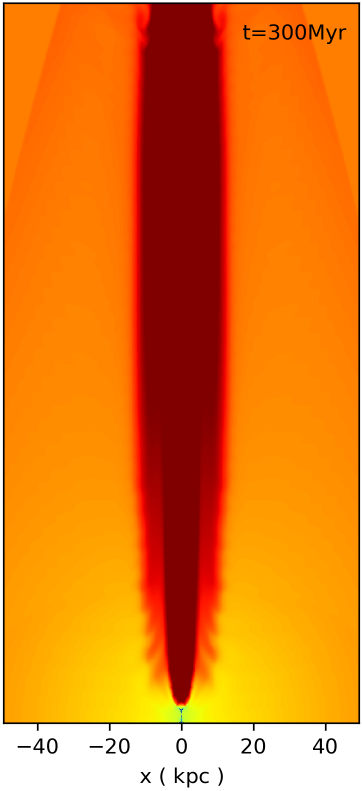}
\includegraphics[height=0.316\textheight]{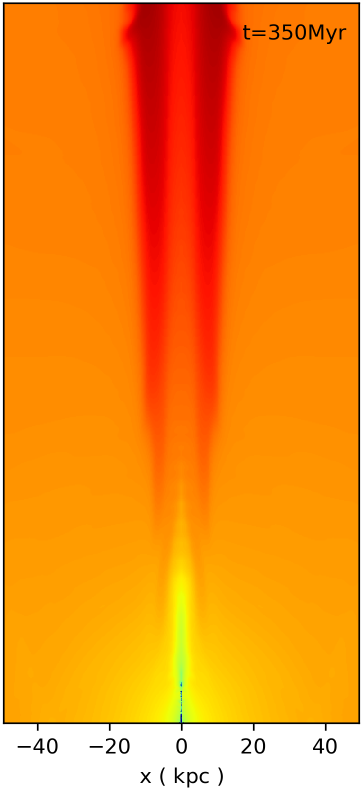}
\includegraphics[height=0.316\textheight]{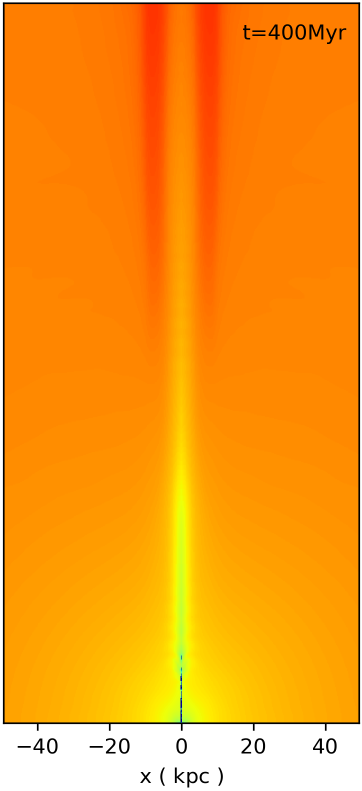}
\includegraphics[height=0.319\textheight]{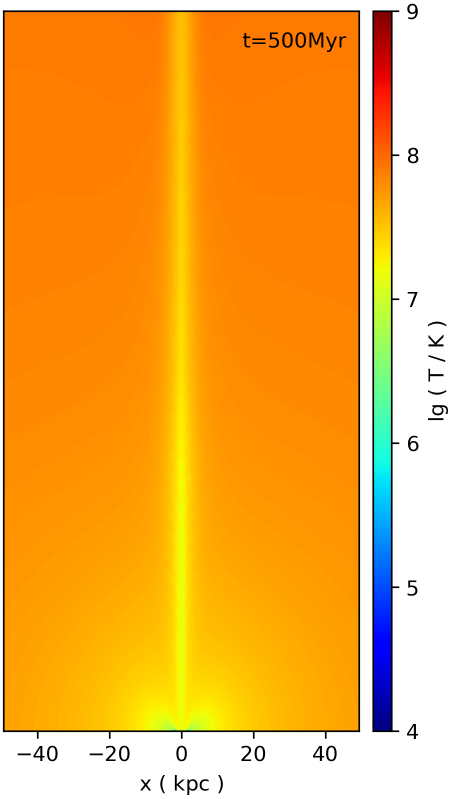}
}
\gridline{
\includegraphics[height=0.340\textheight]{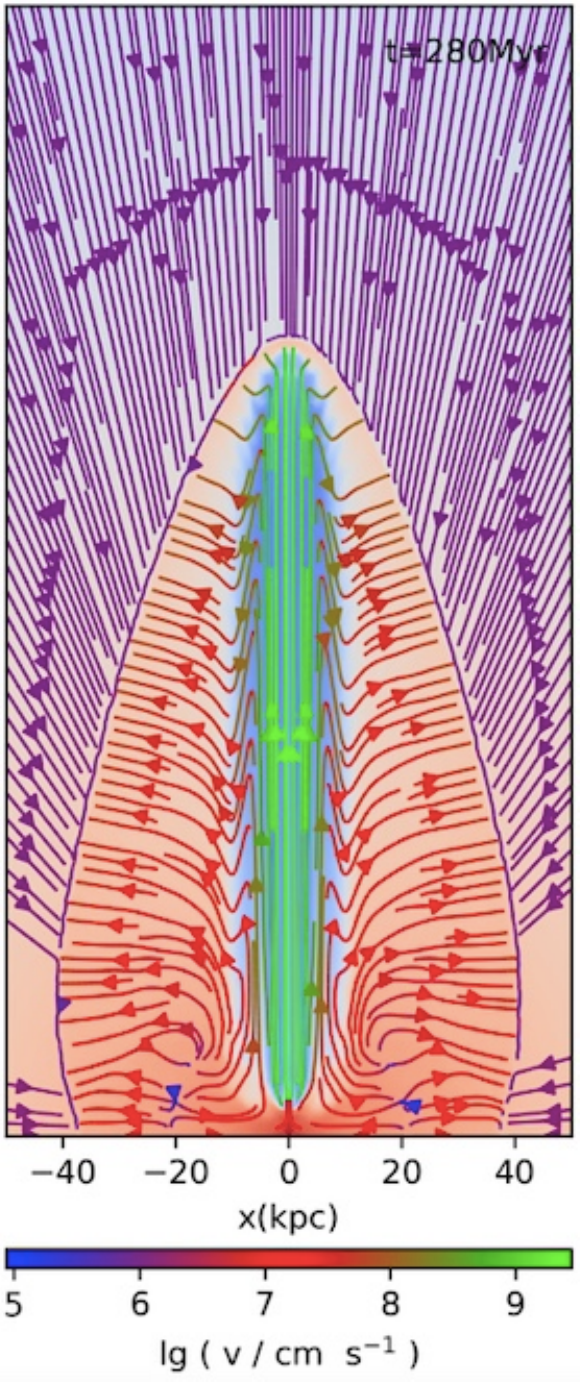}
\includegraphics[height=0.340\textheight]{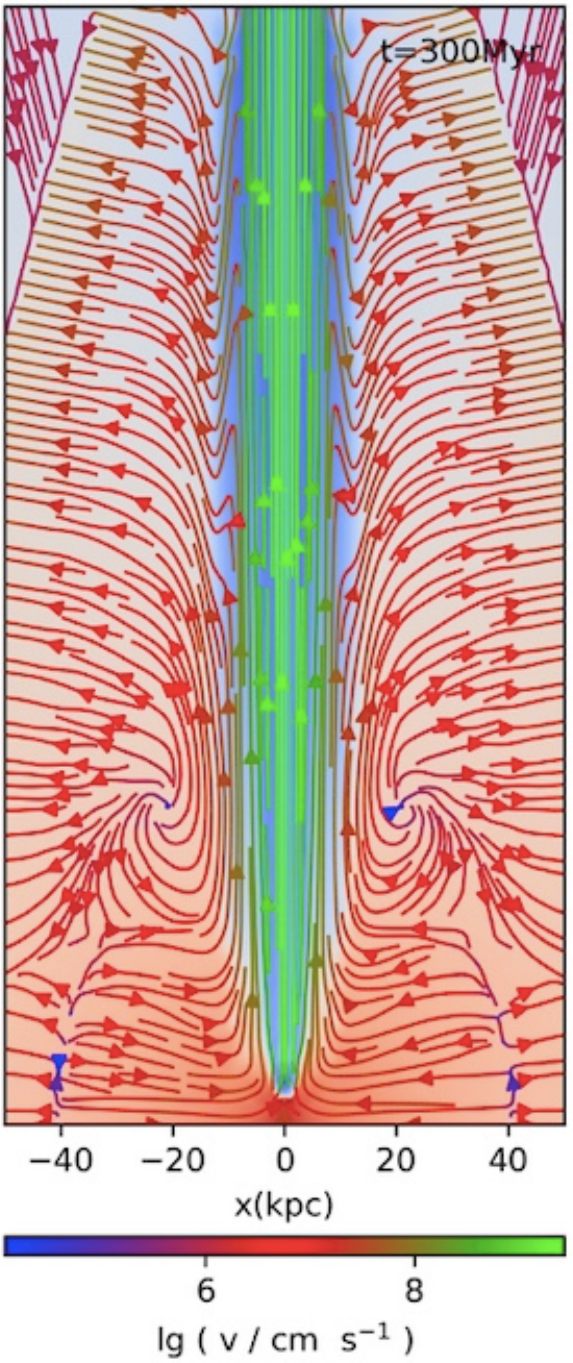}
\includegraphics[height=0.340\textheight]{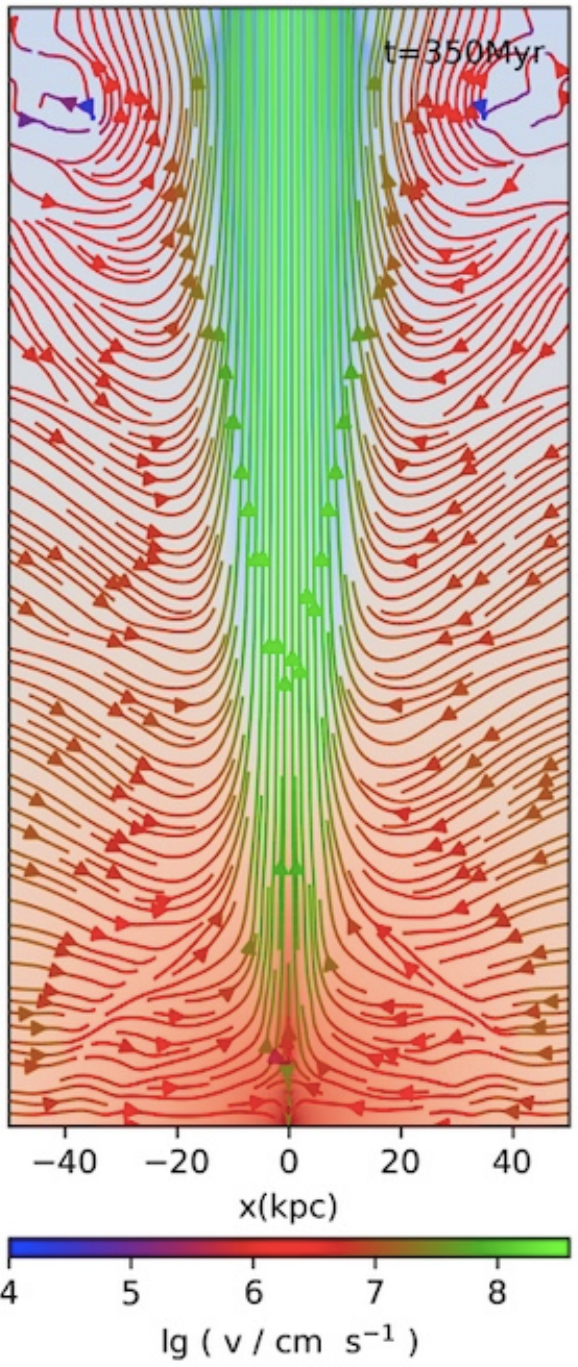}
\includegraphics[height=0.340\textheight]{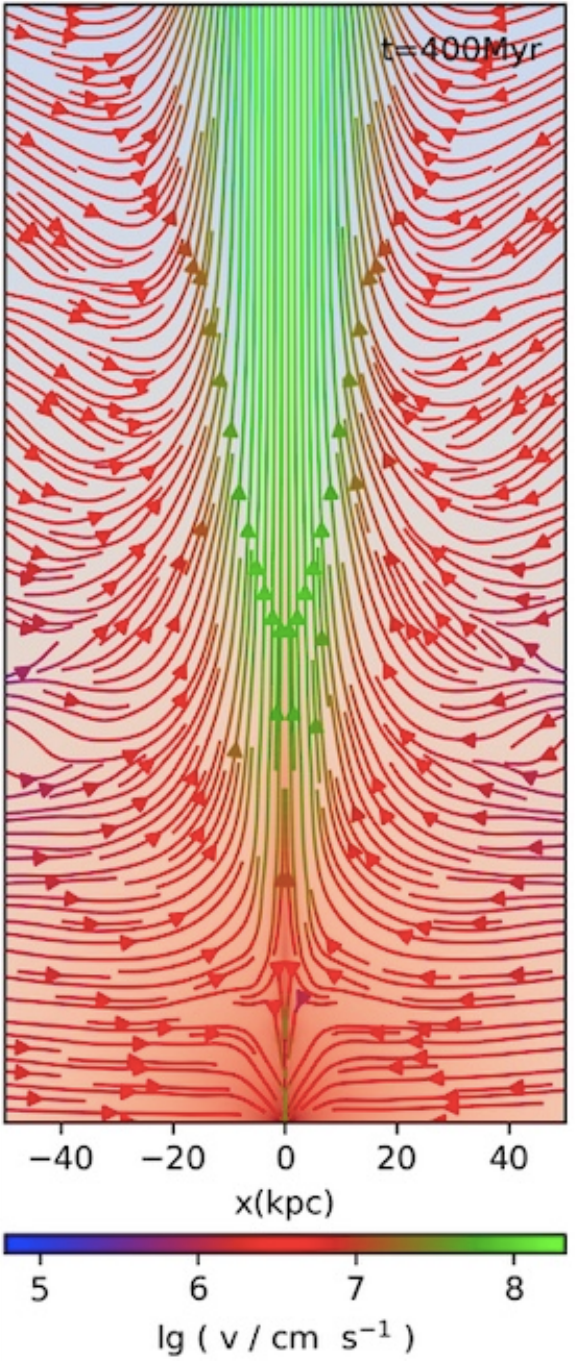}
\includegraphics[height=0.343\textheight]{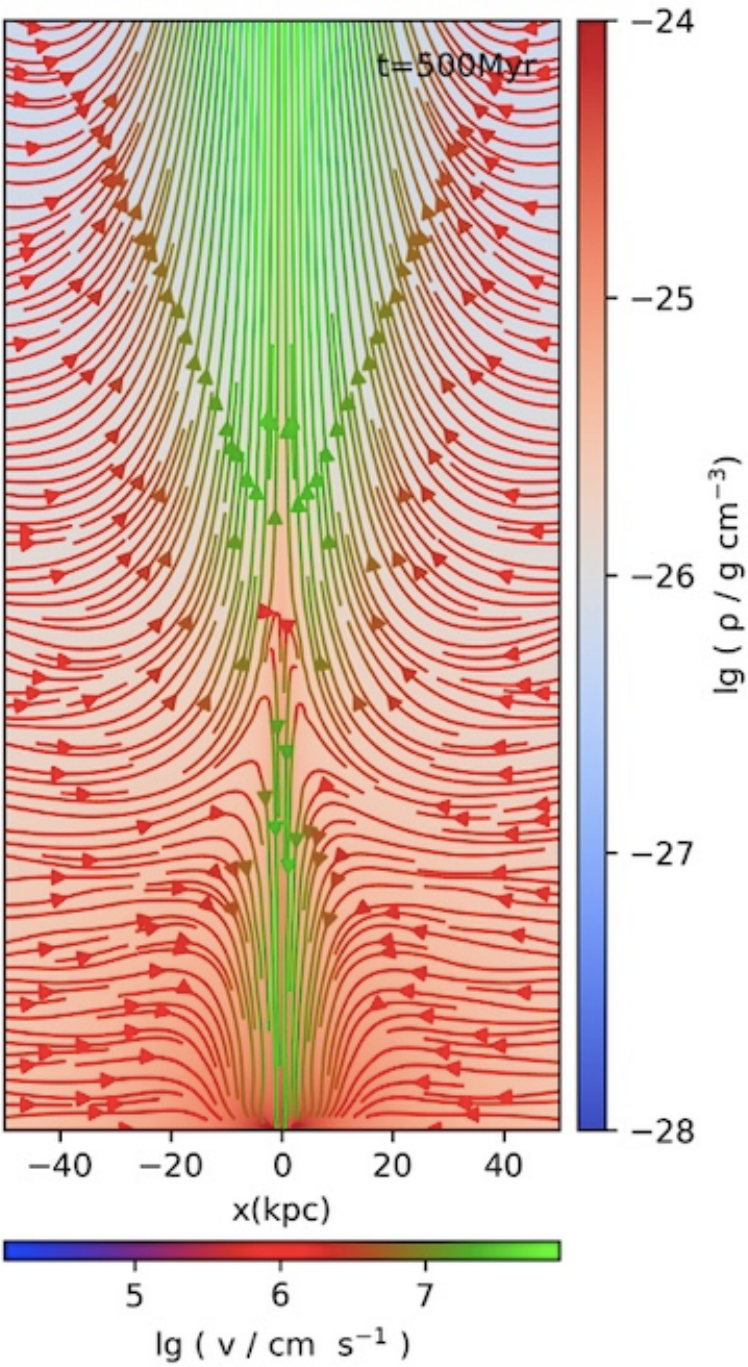}
}
\caption{Same as Figure \ref{figtv-fth9}, but for run Jfth1 with a kinetic-energy-dominated jet.}
\label{figtv-fth1}
\end{figure*}

Assuming axisymmetry around the jet axis, we solve the basic hydrodynamic equations (see \citealt{guo18}) with radiative cooling and gravity in (R, z) cylindrical coordinates using the AMR code MPI-AMRVAC 2.0 (\citealt{xia18}). We choose the galaxy cluster A1795 as our model cluster, following \cite{guo18} to set up the static gravitational potential and the initial conditions. The ICM is initially in hydrostatic equilibrium in the gravitational potential well contributed by a dark matter halo, a central galaxy, and a supermassive black hole. The initial density and temperature distributions provide a good fit to X-ray observations of this cluster (\citealt{guo18}). We adopt the TVDLF scheme (\citealt{keppens12}) in combination with the second-order temporal discretization in this study, and a comparison of the results between the TVDLF and HLLC schemes is shown in Appendix \ref{appA}. We use reflective boundary conditions at the inner boundaries and outflow boundary conditions at the outer boundaries.

In our simulations, the radiative cooling rate is $\mathcal{C}=n_{\rm i}n_{\rm e}\Lambda(T,Z)$, where $n_{\rm e}$ is the electron number density, $n_{\rm i}$ is the ion number density, and the cooling function $\Lambda(T,Z)$ is adopted from \cite{sd93} with a fixed metallicity $Z=0.4$. We use the `exact' cooling method (\citealt{townsend09}) in MPI-AMRVAC 2.0 to calculate cooling processes, and adopt a lower temperature floor $T_{\rm floor}=10^{4}$ K.

The computational domain has a base grid with a resolution of 2 kpc from the origin to 800 kpc and is refined initially with a spatial resolution of 0.25 kpc from the origin to 200 kpc, 0.5 kpc from 200 kpc to 300 kpc, 1 kpc from 300 kpc to 400 kpc, and 2 kpc from 400 kpc to 800 kpc. With the base grid marked as level 1 in MPI-AMRVAC, we use up to six refinement levels, and the finest resolution is 62.5 pc. The mass of cold gas condensation converges as we increase the refinement level to six. Following \cite{li12} and \cite{qiu19} with modifications, an additional grid refinement is triggered to divide a cell into two equal parts when both of the following criteria are satisfied:

(i) The density criterion: the density in this cell meets $\rho \ge 10^{\frac{l}{2}} \rho_{1}$ , where $l$ is the current refinement level, and we choose $\rho_{1} = 3.0 \times 10^{-26}\ \rm g\ cm^{-3}$.

(ii) The cooling time criterion: the cooling time scale in this cell meets $\rm  t_{cool} \le \beta_{cool} t_{cs} $. Here, $\rm t_{cool} = e_{th}/\mathcal{C} $ is the cooling time, where $\rm e_{th}$ is the thermal energy density in the cell and $\mathcal{C}$ is the local cooling rate, and $\rm t_{\rm cs} = \Delta x/c_s$ is the local sound crossing time, where $\rm c_s$ is the local sound speed and $\rm \Delta x$ is the grid size. We choose the factor $\rm \beta_{cool} = 50$ as the mass of cold gas condensation converges when we increase $\rm \beta_{cool} $ to this value.  

 In all our simulations, the ICM first evolves under gravity and radiative cooling. In the pure cooling run without AGN feedback, a cooling catastrophe develops in the ICM at the cluster center at $\rm t = 260$ Myr. In our AGN simulations, we turn on AGN jets at $\rm t =  t_{\rm jeton}$ for a duration of $t_{\rm jet}$. During the AGN phase $\rm t_{\rm jeton} \leq t \leq t_{\rm jeton} + t_{\rm jet}$, a constant jet is injected along the $+z$ direction by adding mass, momentum, and energy into a cylinder with a cross-section radius of $\rm R_{\rm inj} =1$ kpc and a spatial interval of $\rm z_{\rm jet} \le z \le z_{\rm jet} + 1 \rm kpc$ along the $+z$ direction. The injected jet corresponds to a uniform jet with a fixed velocity $\rm v_{\rm jet}=0.1c$, where $c$ is the speed of light. The total energy of one jet injected into our simulated domain is $\rm E_{\rm inj}= 2.36 \times 10^{60} \rm erg$ with a thermal fraction of $\rm f_{\rm th} = E_{\rm th}/ E_{\rm inj}$, where $\rm E_{\rm th}$ is the thermal energy injected by the jet. We choose $\rm f_{\rm th}= 0.9$ and $\rm f_{\rm th}= 0.1$ in our two fiducial runs Jfth9 and Jfth1, respectively, to represent the thermal-energy-dominant jet and the kinetic-energy-dominant jet. In runs Jfth9 and Jfth1, the jet turn-on time is set to be $\rm t_{\rm jeton} = 250$ Myr, just about 10 Myr earlier than the time of cooling catastrophe in our pure cooling run. The jet duration is set to be $\rm t_{\rm jet} = 50$ Myr, which roughly equals to the sound crossing timescale within $50$ kpc in the cluster center and represents the typical mild AGN outbursts as studied in \cite{duan20}. The position of the jet base is set as $\rm z_{\rm jet} = 5$ kpc. The adopted values of the jet parameters in our various AGN runs are shown in Table \ref{tab1}.

\begin{figure*} 
\centering
\gridline{
\includegraphics[height=0.28\textheight]{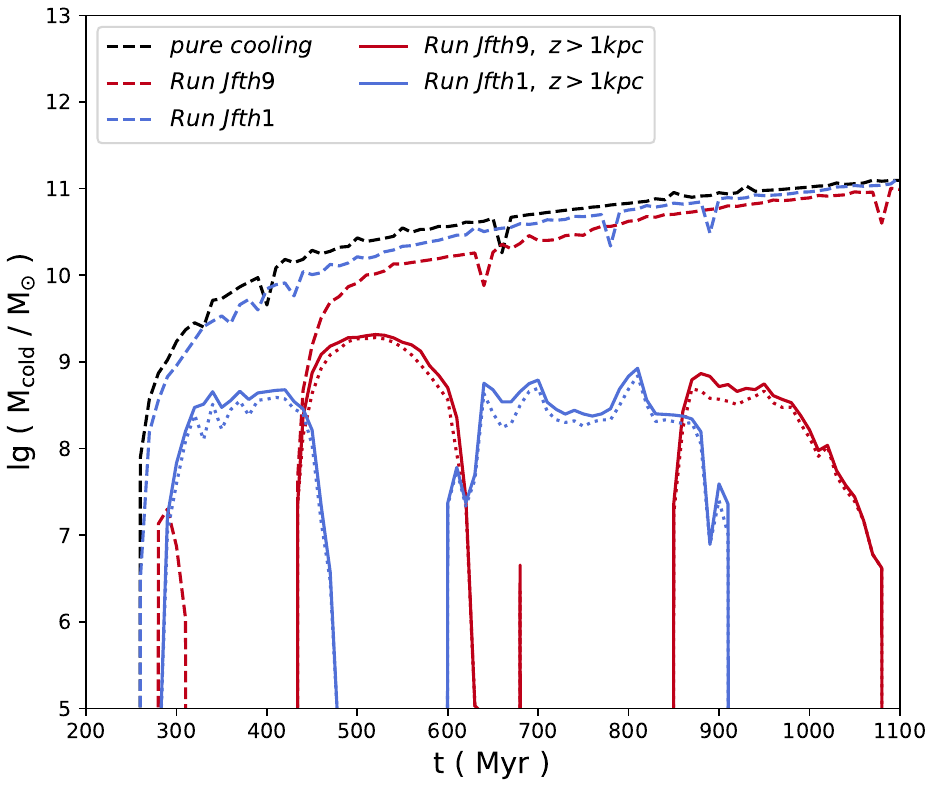}
\includegraphics[height=0.28\textheight]{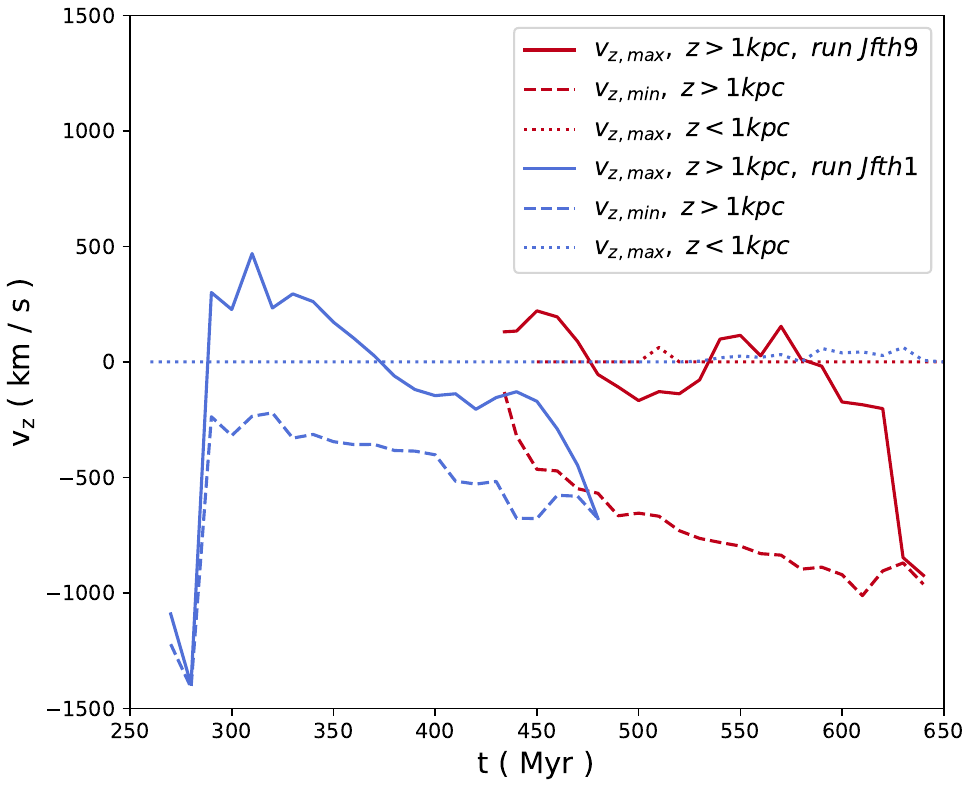}
}
\gridline{
\includegraphics[height=0.28\textheight]{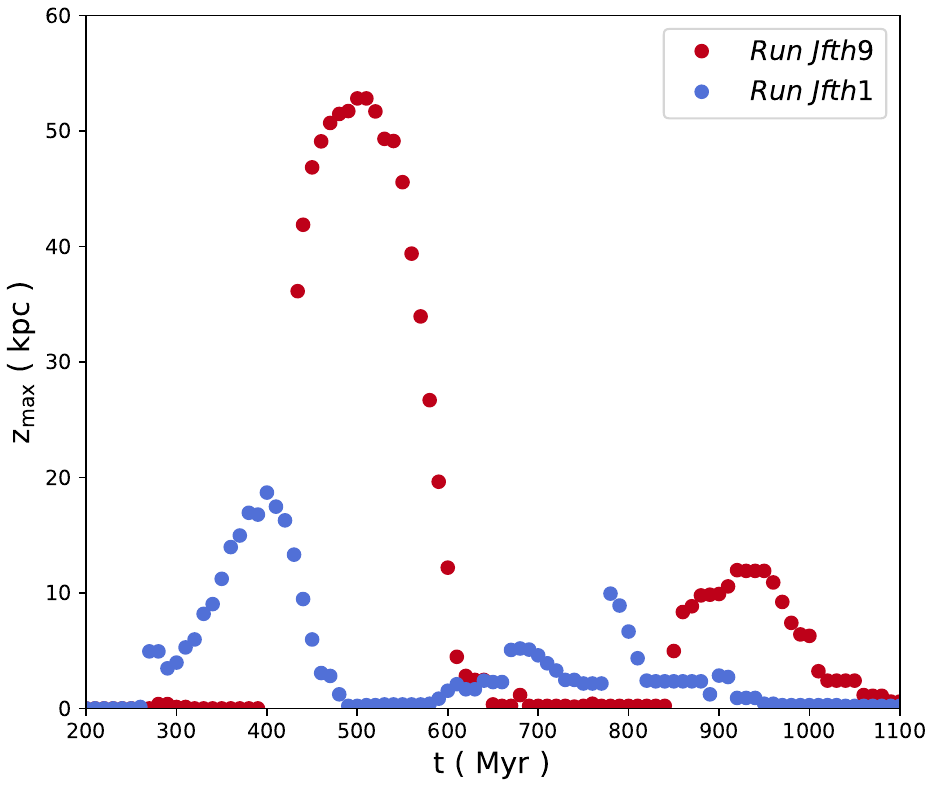}
\includegraphics[height=0.28\textheight]{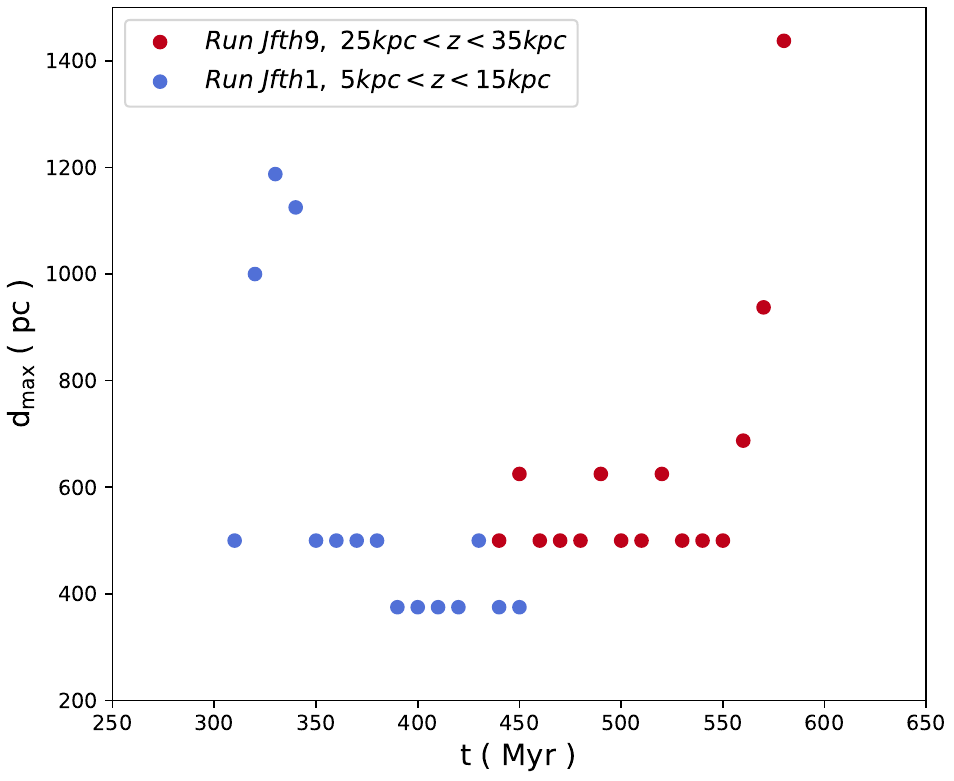}
}
\caption{Temporal evolution of cold gas mass (top left), typical velocity (top right), and spatial distribution (maximum altitude in the z-direction: bottom left, maximum transverse width: bottom right) of cold filaments in runs Jfth9 (red lines and dots) and Jfth1 (blue lines and dots). The cold gas is defined by $T \le T_{\rm cold} = 5 \times 10^5$ K except for the dotted lines in the top left panel where $T_{\rm cold} = 5 \times 10^4$ K. In the top left panel, the black dashed line corresponds to the total mass of cold gas in the pure cooling simulation (with no jetted AGN outburst). The red and blue dashed lines correspond to the total mass of cold gas in runs Jfth9 and Jfth1, respectively, while the solid and dotted lines correspond to the mass of cold gas at $\rm z > 1kpc$, representing the typical mass of cold filaments. In the top right panel, the dashed and solid lines correspond to the minimum and maximum z-component velocities of the cold gas at $\rm z > 1kpc$, and the dotted lines represent the maximum velocity of the cold gas at $\rm z < 1kpc$. In the bottom right panel, $d_{\rm max}$ is defined as the maximum transverse width of the cold filaments located within the spatial intervals of $15$ kpc $< z < 25$ kpc and $5$ kpc $< z < 15$ kpc for runs Jfth9 and Jfth1, respectively.}
\label{figmvzd}
\end{figure*}

\subsection{Morphology of AGN Bubbles and Cold Filaments }
\label{section3.2}

Figure \ref{figtv-fth9} shows the evolution of the gas temperature distribution (top panels) and velocity streamlines (bottom panels; overlapping with the gas density distribution) in run Jfth9 with a thermal-energy-dominated jet ($\rm f_{\rm th}= 0.9$). The jet outburst produces an obvious bow shock, which sweeps the ambient ICM gas outward and heats the gas (\citealt{guo18}; \citealt{duan20}). In run Jfth9, the jet inflates a large bubble at tens of kpc away from the cluster center. The relatively dense gas at the cluster center is then uplifted in the form of a laminar wake flow behind the AGN bubble, as studied in detail in \cite{duan18}. On a larger scale, the wake flow mainly forms within the gas circulations triggered by the anabatic jet-inflated AGN bubbles, as clearly shown in the velocity streamline plots (also see \citealt{guo18} and \citealt{duan18}). At $\rm t = 434\ Myr $, part of the gas in the wake flow cools down to $T \sim 10^{4}$ K, forming short filamentary structures, which then grow to form a long cold filament at about $\rm t = 440\ Myr $. At an earlier time $\rm t = 432\ Myr $, cold gas has not appeared in the wake flow, but the lower part of the wake flow near the cluster center has already begun to fall down toward the cluster center. Thus, we conclude that the cold gas forms by condensation in the hot dense wake flow uplifted by the bubble. As the bubble moves farther away from the cluster center, the cold gas together with a large part of the wake flow continues to fall down to the center. As the wake flow falls down, the gas in the cluster center becomes quite turbulent and may cool down in regions away from the cold filament, as shown in the rightmost panel of Figure \ref{figtv-fth9} ($\rm t = 680 $ Myr). We adopt a lower temperature floor of $\rm T_{\rm floor}=10^{4}$ K in our simulations, and in reality the gas in cold filaments might cool down to a temperature much lower than $10^{4}$ K.

In run Jfth1 with $\rm f_{\rm th}= 0.1$, the kinetic energy-dominated jet cannot inflate a transverse-wide bubble near the cluster center, as shown in Figure \ref{figtv-fth1}. Actually, this kind of mild kinetic energy-dominant jets can only inflate young bubbles elongated along the jet direction (\citealt{perucho19}; \citealt{duan20}; \citealt{guo20}; \citealt{husko23b}), and then evolves into transverse-wide bubbles at a very late stage and far away from the cluster center (\citealt{husko23b}), which may be responsible for giant radio sources (\citealt{dabhade23}). Consequently, the jet in run Jfth1 cannot sustain strong gaseous circulations at the cluster center for a relatively long time, and finally produces a much shorter cold filament than in run Jfth9. Furthermore, unlike in run Jfth9, the cold gas in run Jfth1 appears first near the cluster center, which may be due to the fact that the kinetic energy dominant jet heats the gas at the cluster center less efficiently than the thermal energy dominant jet in run Jfth9 (\citealt{duan20}), as further shown in Section \ref{section3.3}.

\begin{figure*} 
\centering
\gridline{
\includegraphics[height=0.23\textheight]{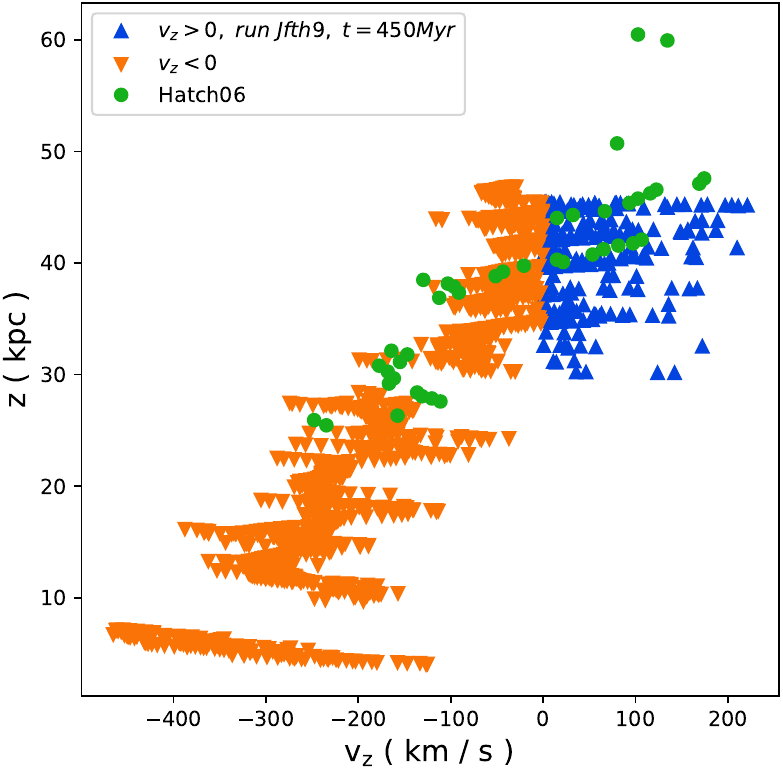}
\includegraphics[height=0.23\textheight]{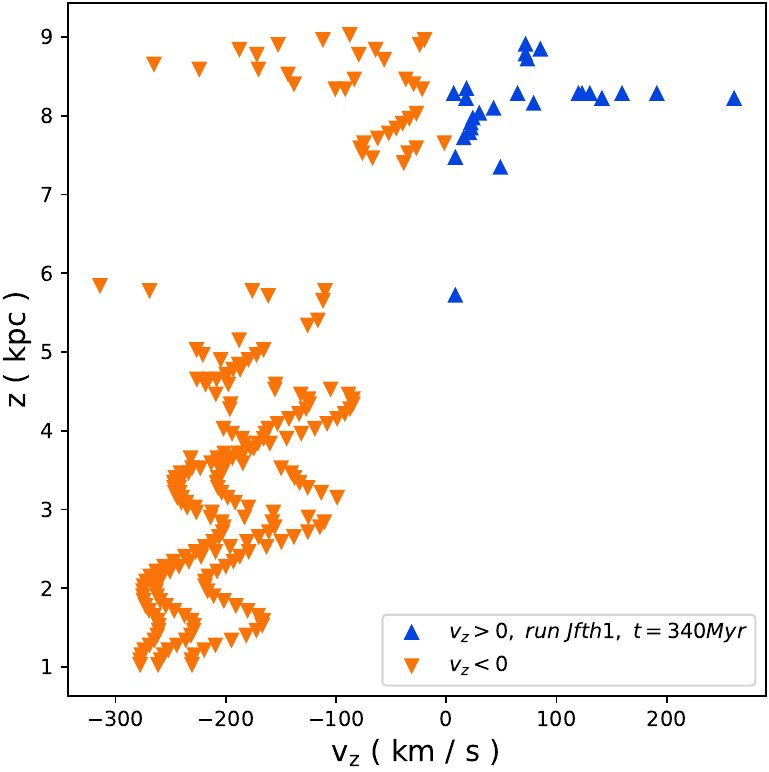}
\includegraphics[height=0.23\textheight]{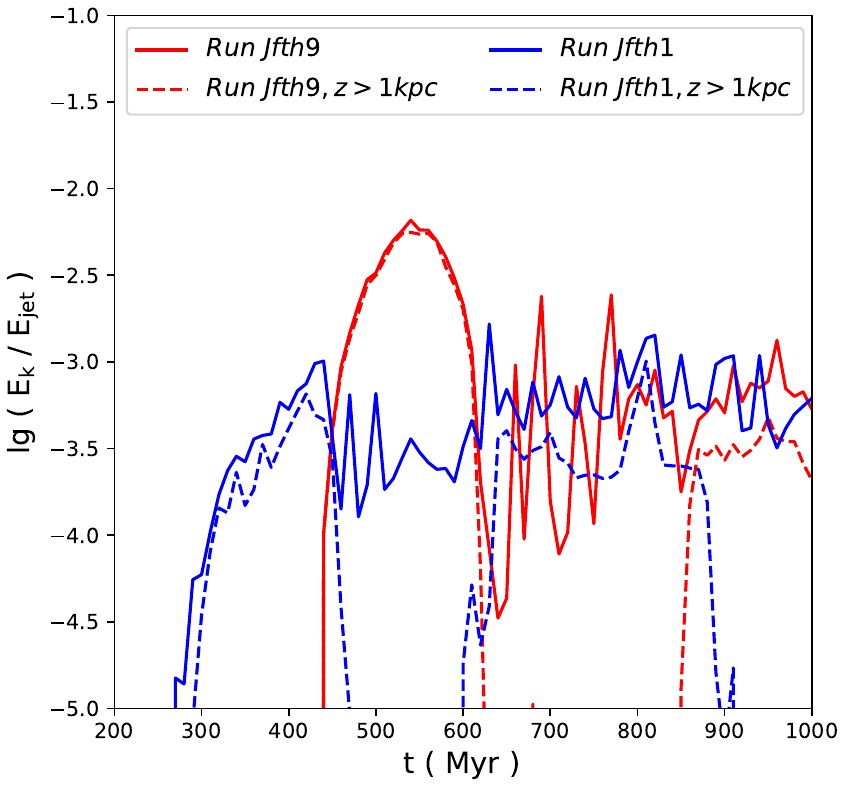}
}
\caption{Left two panels: vertical distributions of the z-component velocity ($\rm v_z$) of the cold gas at $\rm z > 1$ kpc in the cold filaments in runs Jfth9 (left) and Jfth1 (middle) at $t = 450$ Myr (left panel) and $340$ Myr (middle panel), respectively. Each triangle in both panels represents a computational cell at the corresponding time. The positive values of $\rm v_z$ (outflows) are marked with blue triangles, while the negative ones (inflows) are marked with orange triangles. The green circles in the left panel correspond to the observations of the northern cold filament in NGC 1275 shown in Figure 5 of \cite{hatch06}, where line-of-sight velocities and projected distances are measured. Right: temporal evolution of the fraction of the total jet energy converted to the kinetic energy of the total cold gas (solid lines) and cold filaments ($\rm z > 1$ kpc, dashed lines).}
\label{figvz}
\end{figure*}

\subsection{Evolution of cold filaments: mass, distribution and velocity}
\label{section3.3}

We investigate the properties of cold filaments in this subsection. In our analysis, we define cold gas with the gas temperature satisfying $T \le T_{\rm cold} \equiv 5 \times 10^5$ K. Our results are robust to the specific choice of $T_{\rm cold}$. As the gas cooling curve peaks at $T \sim 10^{5.5}$ K (\citealt{schure09}), the gas at around this temperature cools down to $\rm 10^4$ K rapidly, and the amount of gas in the temperature interval $\rm 10^4$ - $\rm 10^6$ K is rare. Figure \ref{figmvzd} shows the evolution of cold gas properties in runs Jfth9 and Jfth1, including the mass, typical velocity, and spatial distribution of cold filaments.

In the pure cooling simulation, the cooling catastrophe in the ICM occurs at $t \sim 260$ Myr when cold gas starts to accumulate in the cluster center. The cold mass accumulation is shown clearly in the top left panel of Figure \ref{figmvzd} (black dashed line). The mass of cold gas in this run increases rapidly after $t \sim 260$ Myr and reaches about $10^{11} \ M_{\odot}$ at $t \sim 950$ Myr. We note that there are several minor dips on the cold gas mass curve due to numerical issues related to the choices of $T_{\rm cold}$ and numerical scheme, which are unimportant for the current study, as further investigated in Appendix \ref{appA}.

The mass and vertical extension of the cold filaments are shown in the left panels of Figure \ref{figmvzd}. In run Jfth9 (red lines and dots), the cooling flow is reversed (also see \citealt{guo18}), and the cooling catastrophe is delayed from $t \sim 260$ Myr in the pure cooling run to $t \sim 430$ Myr in run Jfth9 (red dashed line in the top left panel). During $t \sim 430 - 630$ Myr, a cold filament forms in the wake flow below the AGN bubble (see Figure \ref{figtv-fth9}). As shown in the top-left panel of Figure \ref{figmvzd}, the main part of the cold filament (defined as the cold gas located at $\rm z > 1\ kpc$) has a typical mass of about $\rm 10^9\ M_{\odot}$ (red solid line) during most of its lifetime. The value of this mass is insensitive to the choice of the temperature criterion $T_{\rm cold}$ for cold gas, and a 10-times-lower value of $T_{\rm cold} = 5 \times 10^4 $ K results in a very close mass of the cold filament (red dotted line). The length of the cold filament, defined as the maximum distance $z_{\rm max}$ of the cold gas from the cluster center, reaches $\rm z_{max} \approx 53$ kpc at $\rm t = 500\ Myr$ as shown in the bottom left panel (red dots). 

In run Jfth1 (blue lines and dots), the kinetic-energy-dominated jet penetrates away from the cluster center quickly, resulting in a much less efficient heating on the gas in the cluster central region. Thus, as shown in the top left panel of Figure \ref{figmvzd}, the AGN outburst in run Jfth1 does not delay the cooling catastrophe substantially. As discussed in subsection \ref{section3.2}, a cold filament still forms in this run during $t \sim 290 - 470$ Myr, but with a much lower maximum length of $\rm z_{max} \approx 19$ kpc (at $t \sim 400$ Myr) than in run Jfth9. The cold filament also has a lower typical mass of several $\rm 10^8 M_{\odot}$. As shown in the left panels of Figure \ref{figmvzd}, there is a second short filament in both runs Jfth9 and Jfth1. This feature is not the focus of the current study, and in Appendix \ref{appB}, we show that it is mainly caused by gas motions after the wake flow falls back down to the cluster center at later times.

To investigate the typical width of the simulated cold filaments, we show in the the bottom right panel of Figure \ref{figmvzd} the temporal evolution of the maximum transverse width of the cold filaments located within the spatial intervals of $15$ kpc $< z < 25$ kpc and $5$ kpc $< z < 15$ kpc for runs Jfth9 and Jfth1, respectively. The typical width of the simulated cold filaments is about $500$ pc, especially for the long filament in run Jfth9 (red dots). This result is consistent with some previous observations of cold filaments with typical widths lower than 1 kpc (\citealt{conselice01}; \citealt{hatch06}; \citealt{mcdonald09}; \citealt{fabian16}). There may exist a mixing layer between a cold filament and its surrounding hot gas, where subtle physics such as gas mixing, thermal conduction, and magnetic fields may interplay with the gas cooling process (\citealt{gronke20}; \citealt{nelson20}; \citealt{faucher23}), potentially affecting the width of the cold filament. We leave these detailed microscopic processes for future studies.

To investigate typical gas velocities in the cold filaments, we find the maximum and minimum $z$-component velocities of the cold gas located at $\rm z > 1 $ kpc, denoted as $v_{\rm z,max}$ and $v_{\rm z,min}$, respectively. As shown in the top right panel of Figure \ref{figmvzd}, $v_{\rm z,min}$ usually refers to the maximum infall velocity in the cold filaments, while $v_{\rm z,max}$ refers to the minimum infall velocity or the maximum outflow velocity if some cold gas is outflowing. The maximum infall velocity in our simulated cold filaments is typically a few hundred km/s. At early times, some cold gas may be outflowing, while at later times all the cold gas falls toward the cluster center. The dotted lines in the top right panel of Figure \ref{figmvzd} show the temporal evolution of the maximum vertical velocity at $\rm z < 1$ kpc in runs Jfth1 and Jfth9. In both runs, the maximum vertical velocity of the cold gas in the inner 1 kpc remains zero at early times, indicating that the cold gas appearing at high altitudes does not originate from uplifting the cold gas at the cluster center, but actually forms at $\rm z > 1 $ kpc by cooling out directly from the uplifted dense hot wake flow.

\subsection{Kinetic properties of cold filaments}
\label{section3.4}

In this subsection, we investigate the kinetic properties of the cold filaments in our simulations. The vertical distributions of the z-component velocity ($\rm v_z$) of the cold gas at $\rm z > 1$ kpc in the cold filaments are shown in the left two panels of Figure \ref{figvz}. In particular, we show the $v_{\rm z}-z$ plots of runs Jfth9 and Jfth1 at $t = 450$ Myr and $340$ Myr, respectively. Each triangle in both panels represents a computational cell containing cold gas at $\rm z > 1$ kpc at the corresponding time. Clearly, both cold gas outflows and inflows appear in the filaments at these two representative times (also see the $\rm v_{z} - t $ plot in Figure \ref{figmvzd}). In particular, outflows and inflows coexist at the far side of the filaments from the cluster center, and the inflow velocity then gradually increases toward the cluster center, showing a stretching structure with smooth velocity gradients consistent with the observations of some $\rm H_{\alpha}$ filaments (\citealt{hatch06}; \citealt{mcdonald12}; \citealt{hamer16}). 

In the left panel of Figure \ref{figvz}, we also show the observed line-of-sight velocities and the corresponding projected distances of the northern cold filament in NGC 1275 \citep{hatch06}, and one can see that the results of our fiducial run Jfth9 are roughly consistent with this observation. The kinetic-energy-dominated jet in run Jfth1 could not produce such a long cold filament as observed in NGC 1275. Our simulations further predict that the typical velocity dispersions in cold filaments are several hundred $\rm km\ s^{-1}$. In the rightmost panel of Figure \ref{figvz}, we show the temporal evolution of the fraction of the jet energy converted to the kinetic energy of the cold filaments, which is always less than one percent and consistent with observations (e.g., \citealt{tamhane22}). 

\begin{figure} 
\gridline{
\includegraphics[height=0.28\textheight]{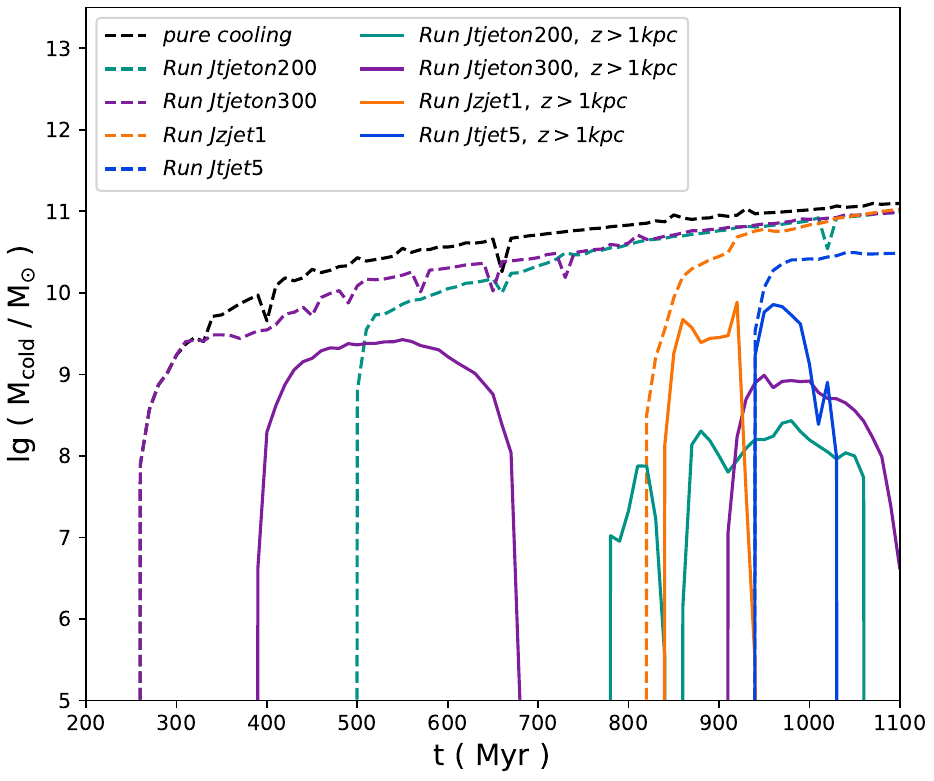}
}\gridline{
\includegraphics[height=0.28\textheight]{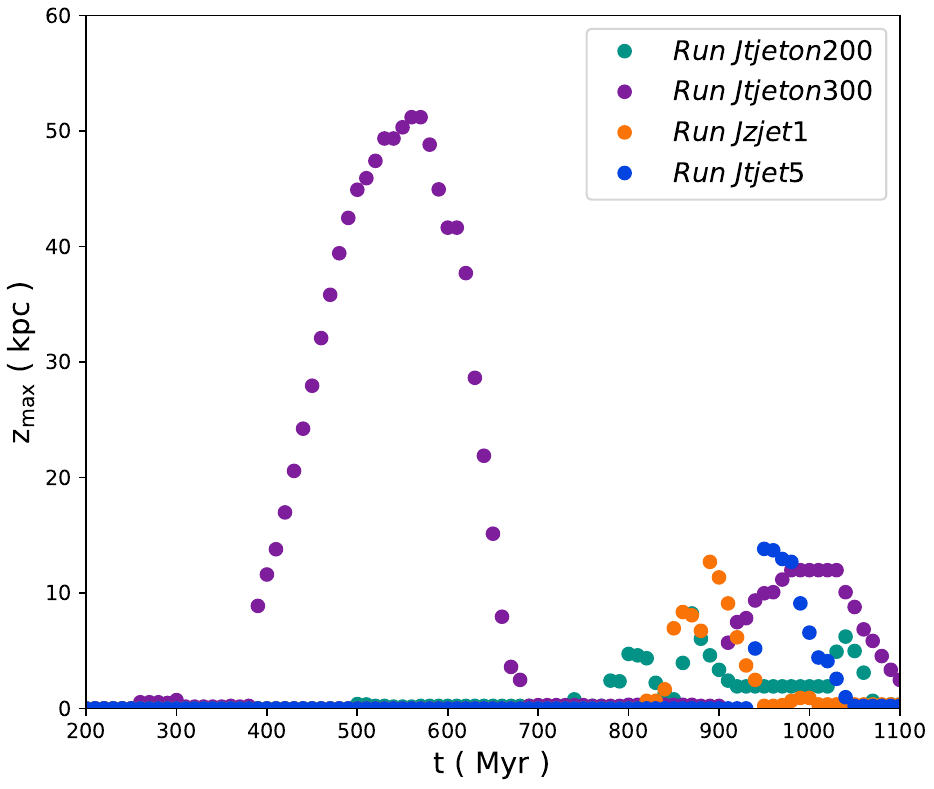}
}
\caption{Same as the left panels of Figure 3, but for runs Jtjeton200, Jtjeton300, Jzjet1, and Jtjet5 with different jet parameters (see Table \ref{tab1} and text in Section \ref{section3.5} for details). }
\label{figmz2}
\end{figure}

\begin{figure*} 
\centering
\includegraphics[height=0.28\textheight]{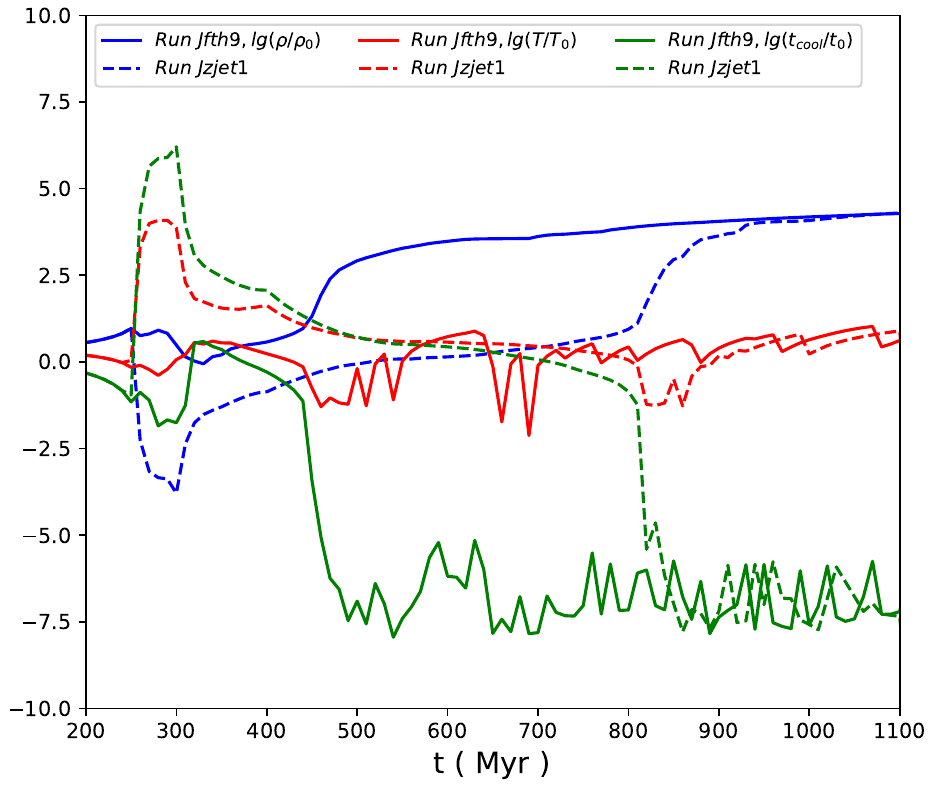}
\includegraphics[height=0.28\textheight]{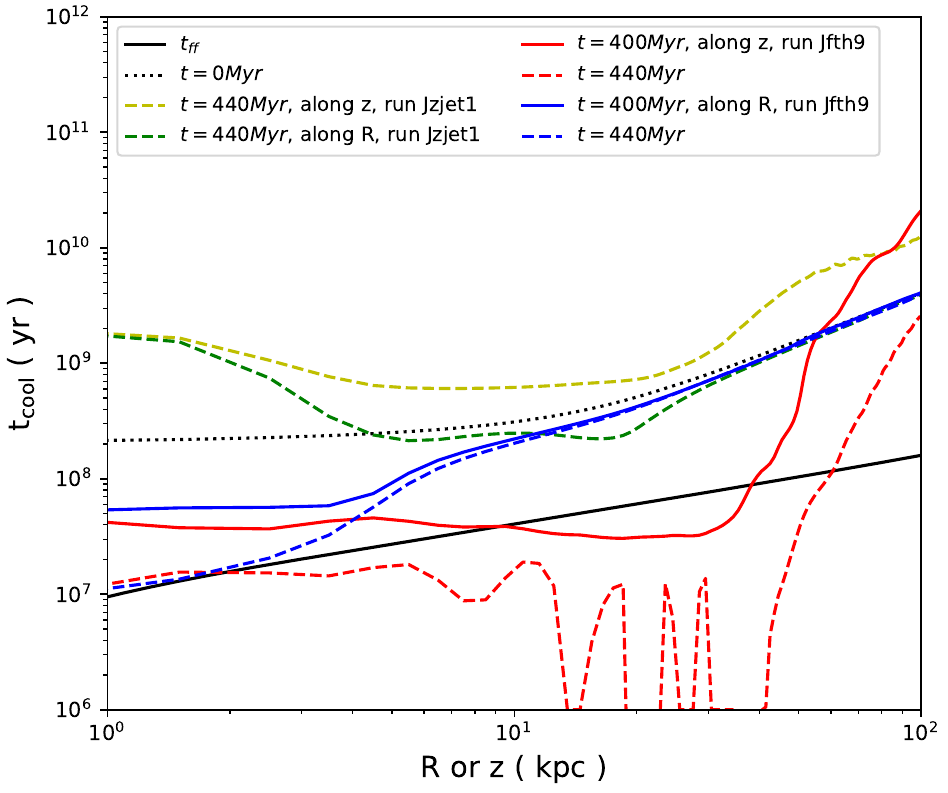}
\caption{Comparisons between runs Jfth9 and Jzjet1. Left: Temporal evolutions of the volume-average gas density (blue lines), the mass-weighted average gas temperature (red lines), and the average gas cooling time (green lines) in the logarithmic scale in the cluster center ($r \leq 1$ kpc) in runs Jfth9 (solid lines) and Jzjet1 (dashed lines). These values are normalized by $\rm \rho_{0} = 10^{-25}\ g\ cm^{-3}$, $\rm T_{0} = 10^{7}\ K$, and $\rm t_{0} = 10^{8}\ yr$, respectively. The average gas cooling time is calculated by dividing the total thermal energy within the $r \leq 1$ kpc region by the corresponding total radiative cooling rate. Right: Radial profiles of the gas cooling time along the $R$ and $z$ axes in runs Jfth9 (red and blue lines) and Jzjet1 (yellow and green lines). The initial gas cooling time (dotted black line) and the freefall time (solid black line) are also shown for comparison.}
\label{fig6}
\end{figure*}

\subsection{Dependence on additional jet parameters}
\label{section3.5}

In addition to the jet's energy content (thermal versus kinetic energies) discussed above, the properties of cold filaments may also be affected by other jet parameters. Here we adopt run Jfth9 as our fiducial run and investigate the impact of additional jet parameters by varying only one additional jet parameter in each comparison run. The parameters of these comparison runs are shown in Table \ref{tab1} and the main results are shown in Figure \ref{figmz2}, as described in detail below.

First, we study the impact of the jet turn-on time, $\rm t_{jeton}$. In run Jfth9, the jet is turned on at $\rm t_{jeton}=$ 250 Myr, slightly before the starting time of the central cooling catastrophe ($t=260$ Myr) in the pure cooling run. Here we consider two additional cases: (i) In run Jtjeton200 with $\rm t_{jeton} = 200$ Myr (green lines in Figure \ref{figmz2}), the cooling catastrophe is delayed to $t=500$ Myr, much later than the delayed starting time of the cooling catastrophe ($t=430$ Myr) in run Jfth9. Before turning on the AGN jet outburst, the ICM gas in the cluster center continues to acquire higher densities and lower entropy due to radiative cooling. Thus, the gas in the cluster center at $\rm t = 200$ Myr has much lower densities and higher entropy than at $\rm t = 250$ Myr in run Jfth9, and consequently after the same AGN outburst, the central lower-density higher-entropy gas in run Jtjeton200 has a much larger cooling time than in run Jfth9. The same argument also explains the significantly-delayed formation time ($\rm t = 760$ Myr) of a much shorter lower-mass ($\sim 10^{8} M_{\odot}$) cold filament in run Jtjeton200 than in our fiducial run. (ii) In run Jtjeton300 with $\rm t_{jeton} = $ 300 Myr (purple lines in Figure \ref{figmz2}), the AGN jet ourburst turned on 40 Myr after the onset of the cooling catastrophe significantly impedes the continuous condensation of cold gas, but can not destroy the cold gas core in the cluster center. Obviously it is hard for our jet outburst initiated at $z=5$ kpc to destroy the dense core with a mass $\sim 10^{9} M_{\odot}$ within 1 kpc from the cluster center. Compared with run Jfth9, the cold filament in run Jtjeton300 with a later jet turn-on time has a comparable mass and maximum length but a longer lifetime.

Second, we study the impact of the location of the jet base $\rm z_{jet}$. To sufficiently resolve the cross section of AGN jets, galaxy-scale simulations often initiate AGN jets at the jet base with a jet radius of order 1 kpc, much larger than in observations (e.g., \citealt{omma04}; \citealt{vernaleo06}; \citealt{gaspari12}; \citealt{weinberger18}; \citealt{blandford19}). To avoid overheating the cluster center by the wide jet, we place the jet base at $\rm z_{jet}$ = 5 kpc in our fiducial run (run Jfth9). To investigate the impact of $\rm z_{jet}$ on our results, here we change the jet base to $\rm z_{jet}$ = 1 kpc in run Jzjet1, much closer to the cluster center than in our fiducial setup. As shown in Figure \ref{figmz2} (orange lines), it is clear that this $\rm z_{jet}$ = 1 kpc jet delays the onset of the cooling catastrophe to $t=820$ Myr, much later than in the fiducial run (Jfth9). The temporal evolutions of the average gas density, temperature, and cooling time within the central 1 kpc of the cluster are shown in the left panel of Figure \ref{fig6}. In run Jzjet1, the spikes at $t\sim 250 - 320$ Myr in the profiles of the temperature and cooling time, and the corresponding density dip are caused by the occupying of the cluster center ($r \leq 1$ kpc) by the jet ejecta, which expel the ICM gas to larger radii. During this process, the ICM gas in the cluster center is overheated and consequently has a long cooling time even when it is dragged out in the wake of the AGN bubble (see the yellow dashed line in the right panel of Fig. \ref{fig6}). Therefore, run Jzjet1 with a lower jet base produces a much shorter cold filament (about 10 kpc) at much later times than our fiducial run. Our result indicates that the location of the jet base significantly affects gas heating and subsequent condensation in the cluster center, suggesting that the heating effect of AGN jet feedback in the cluster center is very subtle and AGN feedback seems to maintain a delicate balance between the ICM cooling and heating.

Finally, we study the impact of the jet power in run Jtjet5. The same amount of jet energy as in run Jfth9 is injected for a duration of 5 Myr in a comparison run Jtjet5, so the jet has a power 10 times stronger. As shown previously in \citet{guo18} and \citet{duan20}, this kind of powerful thermal-energy-dominated jet can heat the cluster center very efficiently and suppress cooling flows very effectively. As shown in Figure \ref{figmz2} (blue lines), this powerful jet delays the onset of the cooling catastrophe to $t=940$ Myr, much later than in the fiducial run (Jfth9). Similar to runs Jtjeton200 and Jzjet1 discussed above, this powerful jet produces a much shorter cold filament at later times than in our fiducial run Jfth9.

Through these results, one can see the properties of cold filaments, especially the filament length, indeed vary in runs with different jet parameters. Long filaments could be produced when the jets are triggered near or after the onset of the central cooling catastrophe and have proper injection powers, thermal fractions, and jet bases. With the same amount of jet energy, if the cluster center is overheated due to a low jet base, a short jet duration, or an early jet turn-on time, gas condensation in the wake flow would occur at a late stage and the formed cold filament would be relatively short (as demonstrated in runs Jtjeton200, Jzjet1, and Jtjet5). As shown in the right panel of Fig. \ref{fig6}, gas condensation in the wake flow in the fiducial run Jfth9 could be explained by the relatively shorter cooling times there compared to those in other regions, which are due to the combined effect of both the uplifting of the central low-entropy high-density gas in the wake flow and the converging motion associated with the uplifting process. Although AGN jet feedback causes the formation of linear wake flows and gas condensation in cold filaments, it generally has a negative-feedback effect on the ICM by heating it and delaying the onset of the central cooling catastrophe.

\section{Summary and Brief Discussion}
\label{section4}

We present a systematic study on the formation of cold filaments in the wake flows uplifted by jet-inflated AGN bubbles in the ICM. To get physical intuition, we first estimate the typical mass of wake flows in Section \ref{section2}, and the result is consistent with our previous simulations (\citealt{duan18}) and X-ray observations of metal-rich outflows in galaxy clusters (\citealt{kirp15}). Using hydrodynamic simulations with high spatial resolution achieved by the AMR technique, here we confirm that cold filaments could indeed form by condensation in hot wake flows behind AGN jet-inflated bubbles. In our analysis, we focus on the formation of the jet-inflated bubbles and cold filaments in the ICM produced by AGN jets with different thermal fractions and also investigate the impact of other jet parameters, such as the jet turn-on time, the position of jet base, and the jet power. The main results are summarized as follows:

(i) Cold thin filaments could naturally form by condensation in the hot dense wake flows uplifted by jet-inflated AGN bubbles. Gas condensation in the wake flow is attributed to the short cooling times there, due to the combined effect of both the uplifting of the central low-entropy high-density gas in the wake flow and the converging motion associated with the uplifting process. To form a long cold filament, the cluster center should not be overheated by the preceding AGN jet outburst. AGN jet feedback seems to maintain a delicate balance between the ICM cooling and heating, and its heating effect in the cluster center is very subtle.

(ii) The simulated cold filaments extend to tens of kpc from the cluster center, having a typical width of about 500 pc and a typical mass of $\rm 10^{8}- 10^{9} M_{\odot}$, which is consistent with previous observations of cold filaments in galaxy clusters (\citealt{conselice01}; \citealt{hatch06}; \citealt{salom08}; \citealt{mcdonald09}; \citealt{olivares19}). The cold filaments in our AGN jet simulations also have similar kinetic properties as of the observations of $\rm H_{\alpha}$ filaments (\citealt{hatch06}; \citealt{mcdonald12}; \citealt{hamer16}) and previous simulations of buoyant bubbles (\citealt{revaz08}; \citealt{brighenti15}). The velocity fields in cold filaments show relatively smooth velocity gradients as in stretching structures. At a representative epoch, the outer part of the filaments contains both inflows and outflows, while the inner part is mainly inflowing toward the cluster center. The typical velocity dispersion in the filament is several hundred $\rm km\ s^{-1}$.

(iii) The properties of cold filaments formed in AGN wake flows depend on the energy content (thermal fraction) of AGN jets. Compared to kinetic-energy-dominated jets, thermal-energy-dominated jets are easier to produce long cold filaments with large masses as observed. They tend to inflate wide low-density bubbles, which stay in the cluster inner regions for a relatively long duration and induce strong gas circulations and wake flows.

(iv) The properties of cold filaments, especially the filament length, are also affected by other jet parameters. AGN jets with an early turn-on time, a low jet base, or a very high power tend to overheat the cluster center, resulting in a long delay in the development of the central cooling catastrophe. The uplifted gas in the wake flows thus takes a long time to cool down and the formed cold filaments are also short in length (see Figure \ref{figmz2}).

We note that our results, especially the lifetime and morphology of cold filaments, may be affected quantitatively by some subtle physical processes such as thermal conduction, turbulent mixing, magnetic fields, cosmic rays (\citealt{gronke20}, \citealt{nelson20}, \citealt{faucher23}), the uncertainty in the gas cooling function (\citealt{schure09}), and even numerical methods (\citealt{martizzi19}; \citealt{weinberger23}). Nonetheless, we expect that the main results of our simulations are robust, such as the general formation processes of cold filaments by condensation in hot dense wake flows of AGN bubbles and the impact of various jet parameters. We further suggest that these jet parameters (e.g., the height of the jet base) should be treated carefully in simulations of galaxies and galaxy clusters containing AGN jet feedback, as they affect both the effectiveness of central gas heating and gas condensation in AGN wake flows.                                                        

\acknowledgements 

We thank the anonymous referee for a constructive report. This work was supported by the High Performance Computing Center of Henan Normal University and the high-performance computing resources in the Core Facility for Advanced Research Computing at Shanghai Astronomical Observatory. FG was partially supported by the Excellent Youth Team Project of the Chinese Academy of Sciences (No. YSBR-061), the Shanghai Pilot Program for Basic Research - Chinese Academy of Science, Shanghai Branch (JCYJ-SHFY-2021-013), and the research grant from the China Manned Space Project (the 2nd-stage CSST science project: Probing Galactic Ecosystems with CSST). 


\appendix


\section{Gas Cooling with Different Resolutions and Numerical Schemes} 
\renewcommand\thefigure{\Alph{section}\arabic{figure}}
\setcounter{figure}{0} 
\label{appA}

\begin{figure*}
\centering
\gridline{
\includegraphics[height=0.29\textheight]{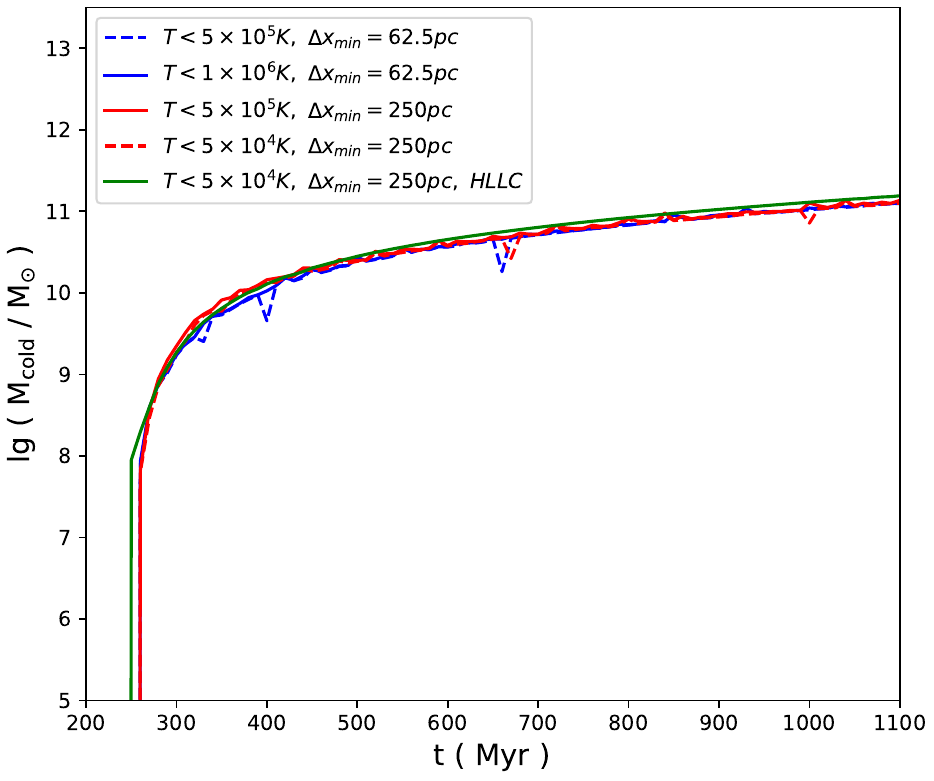}
\includegraphics[height=0.29\textheight]{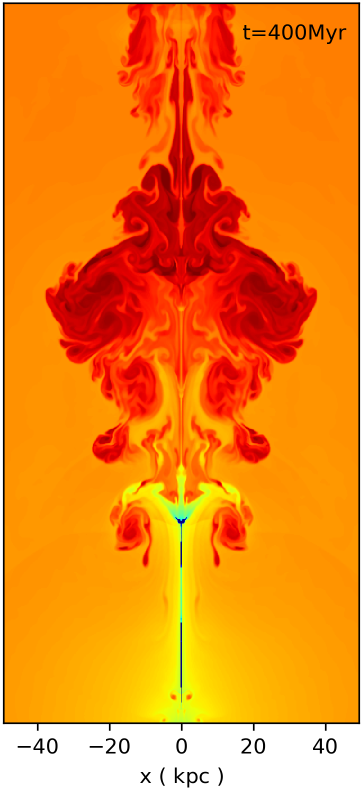}
\includegraphics[height=0.293\textheight]{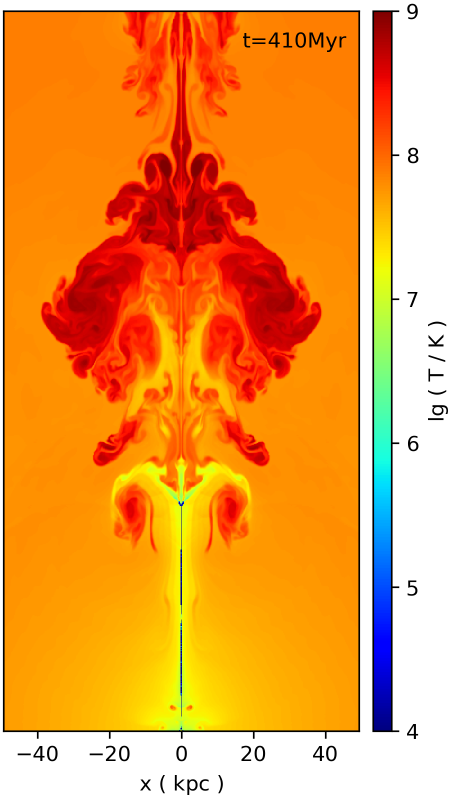}
}
\caption{Comparison of the TVDLF and HLLC schemes. Left: temporal evolution of the cold gas mass in pure cooling runs with different spatial resolutions, cold gas temperature criterion ($T_{\rm cold}$), and numerical schemes. The green solid line represents the result of the run with the HLLC scheme, while the other lines are for TVDLF runs. Middle and right: temperature distributions in a run using the HLLC scheme when the cold filament has just formed ($t=400$, and $410$ Myr). The adopted jet parameters are the same as in the fiducial run Jfth9, except for the jet turn-on time, which is chosen to be 10 Myr earlier as the onset time of the central cooling catastrophe in the pure cooling run with HLLC occurs about 10 Myr earlier.}
\label{figcoolA1}
\end{figure*}

Several minor dips appear on the total cold gas mass curves in Figures \ref{figmvzd} and \ref{figmz2}. We show here that this feature is related to the numerical scheme, spatial resolution and our cold gas criterion, which have rather minor impacts on the overall evolution of AGN bubbles and cold filaments. In the left panel of Figure \ref{figcoolA1}, we present a comparison of cold gas condensation in pure cooling runs using different cold gas temperature criterion ($T_{\rm cold}$), spatial resolutions and numerical schemes. The blue dashed line repeats the black dashed line in the top left panel of Figure \ref{figmvzd} having several minor dips. These dips disappear when we increase $T_{\rm cold}$ from $5\times 10^5$ K to $10^6$ K (blue solid line), or change the finest resolution from $\rm \Delta x_{min} =62.5 pc$ to $\rm 250 pc$ (red solid line). However, if we lower $T_{\rm cold}$ to $5\times 10^4$ K in the low resolution run, several dips appear again (red dashed line). Furthermore, the dips also disappear if the numerical scheme is changed from TVDLF to HLLC (green line). We suggest that the appearance of these minor dips is a manifestation of numerical diffusion when cold gas moves through the hot ICM (\citealt{leveque98}; \citealt{leveque02}). Compared to HLLC, the dips are easier to appear in the more diffusive TVDLF scheme (\citealt{keppens12}). Nonetheless, except for the minor dips that appear only several times on some lines, all the curves are quite close to each other.

To further show the impact of the numerical scheme, we show the morphology of AGN bubbles and cold filaments in the HLLC run in the middle and right panels of Figure \ref{figcoolA1}. Compared to the TVDLF scheme (run Jfth9 in Figure \ref{figtv-fth9}), the formed AGN bubble in the HLLC run has more small turbulent structures due to the Kelvin-Helmholz instability, which are smoothed in the more diffusive TVDLF run similar to the effect of viscosity (\citealt{leveque98}; \citealt{reynolds05}; \citealt{duan18}). The formed cold filament in the HLLC run is also slightly longer than that in the TVDLF run (run Jfth9; see Figures \ref{figtv-fth9} and \ref{figmvzd}). However, the HLLC scheme is too time-consuming for us to conduct these high-resolution cluster-sized AMR simulations with high-velocity flows (jets) and radiative cooling. Instead of investigating the impact of numerical methods in jet feedback simulations (see, e.g., \citealt{martizzi19}; \citealt{weinberger23}), here we choose the TVDLF scheme, which allows us to perform long-duration simulations to investigate the entire formation history of cold filaments and compare the results in various runs with different jet parameters. 

\section{Formation of the Second Short Filament}
\setcounter{figure}{0} 
\label{appB}

\begin{figure} 
\centering
\gridline{
\includegraphics[height=0.29\textheight]{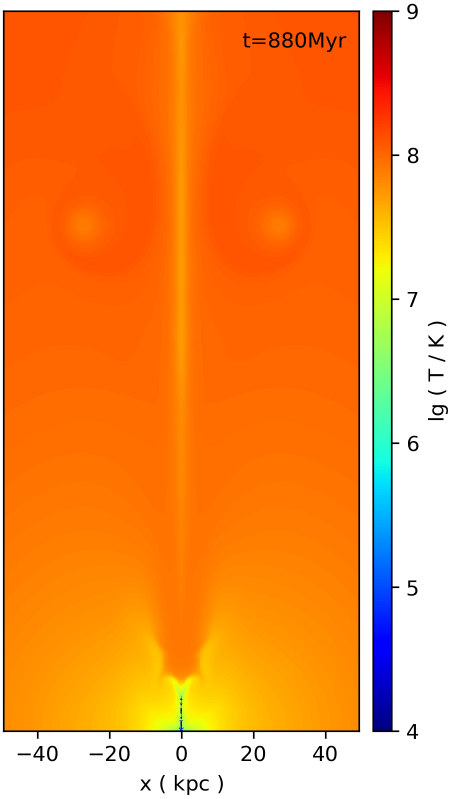}
\includegraphics[height=0.29\textheight]{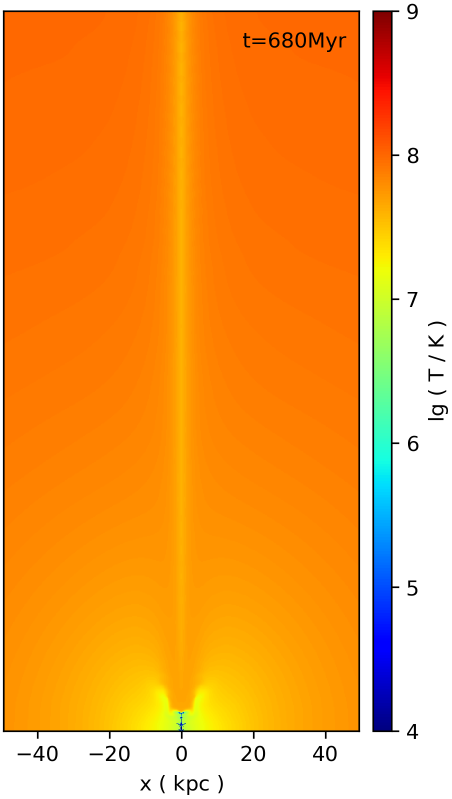}
}
\gridline{
\includegraphics[height=0.295\textheight]{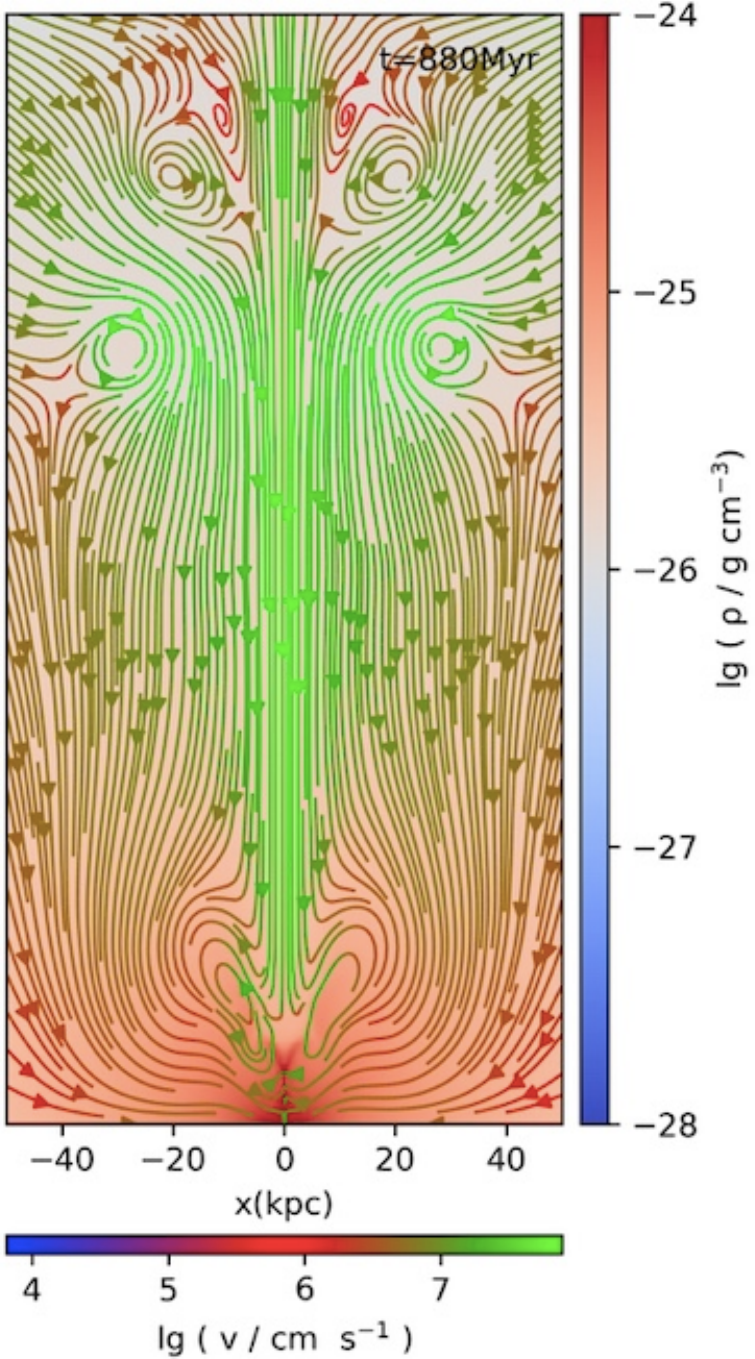}
\includegraphics[height=0.295\textheight]{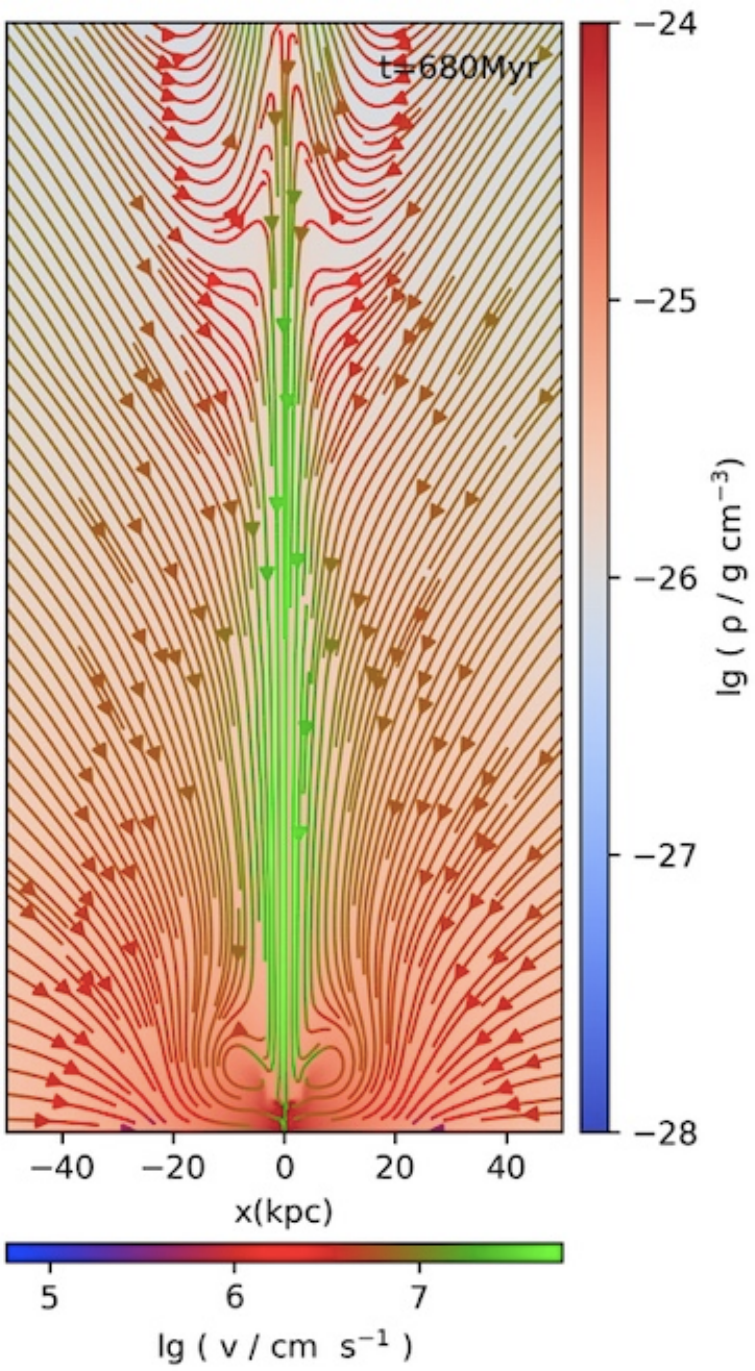}
}
\caption{The same as Figure 1, but showing the temporal evolution of the temperature distributions (top) and gas velocity streamlines overlapping with the density distributions (bottom) in runs Jfth9 (left) and Jfth1 (right) at $t=880$ Myr and $t=680$ Myr, respectively. The second cold filaments appear near the cluster center at these time epochs.  }
\label{figfilament2}
\end{figure}

In Figure \ref{figfilament2}, we show the short filaments near the cluster center in run Jfth9 and run Jfth1 at very late times $t=880$ and $680$ Myr, respectively, corresponding to the second bumps of the mass and length ($z_{\rm max}$) of cold filaments shown in Figure \ref{figmvzd}. From the streamline plots, one can see that this phenomenon is caused by gas motions near the cluster center, long after the major part of the wake flow falls down to the cluster center. This phenomenon also occurs in other runs (see Figure \ref{figmz2}) and in our previous study (\citealt{duan18}; see Figures 4 and 5 therein), and is not the focus of the current study.

\clearpage

\bibliography{ms} 

\begin{thebibliography}{}
\expandafter\ifx\csname natexlab\endcsname\relax\def\natexlab#1{#1}\fi
\providecommand{\url}[1]{\href{#1}{#1}}

\bibitem[{{Beckmann} {et~al.}(2019){Beckmann}, {Dubois}, {Guillard}, {Salome},
  {Olivares}, {Polles}, {Cadiou}, {Combes}, {Hamer}, {Lehnert}, \& {Pineau des
  Forets}}]{beckmann19}
{Beckmann}, R.~S., {Dubois}, Y., {Guillard}, P., {et~al.} 2019, \aap, 631, A60

\bibitem[{{Best} {et~al.}(1996){Best}, {Longair}, \& {Rottgering}}]{best96}
{Best}, P.~N., {Longair}, M.~S., \& {Rottgering}, H.~J.~A. 1996, \mnras, 280,
  L9

\bibitem[{{Blandford} {et~al.}(2019){Blandford}, {Meier}, \&
  {Readhead}}]{blandford19}
{Blandford}, R., {Meier}, D., \& {Readhead}, A. 2019, \araa, 57, 467

\bibitem[{{Boehringer} {et~al.}(1993){Boehringer}, {Voges}, {Fabian}, {Edge},
  \& {Neumann}}]{boehringer93}
{Boehringer}, H., {Voges}, W., {Fabian}, A.~C., {Edge}, A.~C., \& {Neumann},
  D.~M. 1993, \mnras, 264, L25

\bibitem[{{Bourne} \& {Yang}(2023)}]{bourne23}
{Bourne}, M.~A., \& {Yang}, H.-Y.~K. 2023, Galaxies, 11, 73

\bibitem[{{Brighenti} {et~al.}(2015){Brighenti}, {Mathews}, \&
  {Temi}}]{brighenti15}
{Brighenti}, F., {Mathews}, W.~G., \& {Temi}, P. 2015, \apj, 802, 118

\bibitem[{{Br{\"u}ggen}(2003)}]{bruggen03}
{Br{\"u}ggen}, M. 2003, \apj, 592, 839

\bibitem[{{Calzadilla} {et~al.}(2022){Calzadilla}, {McDonald}, {Donahue},
  {McNamara}, {Fogarty}, {Gaspari}, {Gitti}, {Russell}, {Tremblay}, {Voit}, \&
  {Ubertosi}}]{calzadilla22}
{Calzadilla}, M.~S., {McDonald}, M., {Donahue}, M., {et~al.} 2022, \apj, 940,
  140

\bibitem[{{Chambers} {et~al.}(1987){Chambers}, {Miley}, \& {van
  Breugel}}]{chambers87}
{Chambers}, K.~C., {Miley}, G.~K., \& {van Breugel}, W. 1987, \nat, 329, 604

\bibitem[{{Churazov} {et~al.}(2001){Churazov}, {Br{\"u}ggen}, {Kaiser},
  {B{\"o}hringer}, \& {Forman}}]{churazov01}
{Churazov}, E., {Br{\"u}ggen}, M., {Kaiser}, C.~R., {B{\"o}hringer}, H., \&
  {Forman}, W. 2001, \apj, 554, 261

\bibitem[{{Churazov} {et~al.}(2013){Churazov}, {Ruszkowski}, \&
  {Schekochihin}}]{churazov13}
{Churazov}, E., {Ruszkowski}, M., \& {Schekochihin}, A. 2013, \mnras, 436, 526

\bibitem[{{Conselice} {et~al.}(2001){Conselice}, {Gallagher}, \&
  {Wyse}}]{conselice01}
{Conselice}, C.~J., {Gallagher}, John~S., I., \& {Wyse}, R. F.~G. 2001, \aj,
  122, 2281

\bibitem[{{Dabhade} {et~al.}(2023){Dabhade}, {Saikia}, \& {Mahato}}]{dabhade23}
{Dabhade}, P., {Saikia}, D.~J., \& {Mahato}, M. 2023, Journal of Astrophysics
  and Astronomy, 44, 13

\bibitem[{{Dabiri}(2006)}]{dabiri06}
{Dabiri}, J.~O. 2006, Journal of Fluid Mechanics, 547, 105

\bibitem[{{Darwin}(1953)}]{darwin53}
{Darwin}, C. 1953, Proceedings of the Cambridge Philosophical Society, 49, 342

\bibitem[{{Donahue} \& {Voit}(2022)}]{donahue22}
{Donahue}, M., \& {Voit}, G.~M. 2022, \physrep, 973, 1

\bibitem[{{Duan} \& {Guo}(2018)}]{duan18}
{Duan}, X., \& {Guo}, F. 2018, \apj, 861, 106

\bibitem[{{Duan} \& {Guo}(2020)}]{duan20}
---. 2020, \apj, 896, 114

\bibitem[{{Fabian}(2012)}]{fabian12}
{Fabian}, A.~C. 2012, \araa, 50, 455

\bibitem[{{Fabian} {et~al.}(2003){Fabian}, {Sanders}, {Crawford}, {Conselice},
  {Gallagher}, \& {Wyse}}]{fabian03}
{Fabian}, A.~C., {Sanders}, J.~S., {Crawford}, C.~S., {et~al.} 2003, \mnras,
  344, L48

\bibitem[{{Fabian} {et~al.}(2016){Fabian}, {Walker}, {Russell}, {Pinto},
  {Canning}, {Salome}, {Sanders}, {Taylor}, {Zweibel}, {Conselice}, {Combes},
  {Crawford}, {Ferland}, {Gallagher}, {Hatch}, {Johnstone}, \&
  {Reynolds}}]{fabian16}
{Fabian}, A.~C., {Walker}, S.~A., {Russell}, H.~R., {et~al.} 2016, \mnras, 461,
  922

\bibitem[{{Fan} {et~al.}(2018){Fan}, {Wu}, \& {Liao}}]{fan18}
{Fan}, X.-L., {Wu}, Q., \& {Liao}, N.-H. 2018, \apj, 861, 97

\bibitem[{{Faucher-Giguere} \& {Oh}(2023)}]{faucher23}
{Faucher-Giguere}, C.-A., \& {Oh}, S.~P. 2023, arXiv e-prints, arXiv:2301.10253

\bibitem[{{Gaspari} {et~al.}(2012){Gaspari}, {Ruszkowski}, \&
  {Sharma}}]{gaspari12}
{Gaspari}, M., {Ruszkowski}, M., \& {Sharma}, P. 2012, \apj, 746, 94

\bibitem[{{Gronke} \& {Oh}(2020)}]{gronke20}
{Gronke}, M., \& {Oh}, S.~P. 2020, \mnras, 492, 1970

\bibitem[{{Guo}(2015)}]{guo15}
{Guo}, F. 2015, \apj, 803, 48

\bibitem[{{Guo}(2020)}]{guo20}
---. 2020, \apj, 903, 3

\bibitem[{{Guo} {et~al.}(2018){Guo}, {Duan}, \& {Yuan}}]{guo18}
{Guo}, F., {Duan}, X., \& {Yuan}, Y.-F. 2018, \mnras, 473, 1332

\bibitem[{{Hamer} {et~al.}(2016){Hamer}, {Edge}, {Swinbank}, {Wilman},
  {Combes}, {Salom{\'e}}, {Fabian}, {Crawford}, {Russell}, {Hlavacek-Larrondo},
  {McNamara}, \& {Bremer}}]{hamer16}
{Hamer}, S.~L., {Edge}, A.~C., {Swinbank}, A.~M., {et~al.} 2016, \mnras, 460,
  1758

\bibitem[{{Hardcastle} \& {Croston}(2020)}]{hardcastle20}
{Hardcastle}, M.~J., \& {Croston}, J.~H. 2020, \nar, 88, 101539

\bibitem[{{Hatch} {et~al.}(2006){Hatch}, {Crawford}, {Johnstone}, \&
  {Fabian}}]{hatch06}
{Hatch}, N.~A., {Crawford}, C.~S., {Johnstone}, R.~M., \& {Fabian}, A.~C. 2006,
  \mnras, 367, 433

\bibitem[{{Hu{\v{s}}ko} \& {Lacey}(2023{\natexlab{a}})}]{husko23a}
{Hu{\v{s}}ko}, F., \& {Lacey}, C.~G. 2023{\natexlab{a}}, \mnras, 520, 5090

\bibitem[{{Hu{\v{s}}ko} \& {Lacey}(2023{\natexlab{b}})}]{husko23b}
---. 2023{\natexlab{b}}, \mnras, 521, 4375

\bibitem[{{Keppens} {et~al.}(2012){Keppens}, {Meliani}, {van Marle}, {Delmont},
  {Vlasis}, \& {van der Holst}}]{keppens12}
{Keppens}, R., {Meliani}, Z., {van Marle}, A.~J., {et~al.} 2012, Journal of
  Computational Physics, 231, 718

\bibitem[{{Kirkpatrick} \& {McNamara}(2015)}]{kirp15}
{Kirkpatrick}, C.~C., \& {McNamara}, B.~R. 2015, \mnras, 452, 4361

\bibitem[{{Kirkpatrick} {et~al.}(2011){Kirkpatrick}, {McNamara}, \&
  {Cavagnolo}}]{kirp11}
{Kirkpatrick}, C.~C., {McNamara}, B.~R., \& {Cavagnolo}, K.~W. 2011, \apjl,
  731, L23

\bibitem[{LeVeque(1998)}]{leveque98}
LeVeque, R.~J. 1998, Nonlinear Conservation Laws and Finite Volume Methods, ed.
  O.~Steiner \& A.~Gautschy (Berlin, Heidelberg: Springer Berlin Heidelberg),
  1--159

\bibitem[{LeVeque(2002)}]{leveque02}
---. 2002, Finite Volume Methods for Hyperbolic Problems, Cambridge Texts in
  Applied Mathematics (Cambridge University Press)

\bibitem[{{Li} \& {Bryan}(2012)}]{li12}
{Li}, Y., \& {Bryan}, G.~L. 2012, \apj, 747, 26

\bibitem[{{Li} \& {Bryan}(2014)}]{li14}
---. 2014, \apj, 789, 153

\bibitem[{{Martizzi} {et~al.}(2019){Martizzi}, {Quataert},
  {Faucher-Gigu{\`e}re}, \& {Fielding}}]{martizzi19}
{Martizzi}, D., {Quataert}, E., {Faucher-Gigu{\`e}re}, C.-A., \& {Fielding}, D.
  2019, \mnras, 483, 2465

\bibitem[{{McCarthy} {et~al.}(1987){McCarthy}, {van Breugel}, {Spinrad}, \&
  {Djorgovski}}]{mccarthy87}
{McCarthy}, P.~J., {van Breugel}, W., {Spinrad}, H., \& {Djorgovski}, S. 1987,
  \apjl, 321, L29

\bibitem[{{McDonald} \& {Veilleux}(2009)}]{mcdonald09}
{McDonald}, M., \& {Veilleux}, S. 2009, \apjl, 703, L172

\bibitem[{{McDonald} {et~al.}(2012){McDonald}, {Veilleux}, \&
  {Rupke}}]{mcdonald12}
{McDonald}, M., {Veilleux}, S., \& {Rupke}, D. S.~N. 2012, \apj, 746, 153

\bibitem[{{McDonald} {et~al.}(2010){McDonald}, {Veilleux}, {Rupke}, \&
  {Mushotzky}}]{mcdonald10}
{McDonald}, M., {Veilleux}, S., {Rupke}, D. S.~N., \& {Mushotzky}, R. 2010,
  \apj, 721, 1262

\bibitem[{{McNamara} \& {Nulsen}(2012)}]{mcnamara12}
{McNamara}, B.~R., \& {Nulsen}, P.~E.~J. 2012, New Journal of Physics, 14,
  055023

\bibitem[{{McNamara} {et~al.}(2016){McNamara}, {Russell}, {Nulsen}, {Hogan},
  {Fabian}, {Pulido}, \& {Edge}}]{mcnamara16}
{McNamara}, B.~R., {Russell}, H.~R., {Nulsen}, P.~E.~J., {et~al.} 2016, \apj,
  830, 79

\bibitem[{{Nelson} {et~al.}(2020){Nelson}, {Sharma}, {Pillepich}, {Springel},
  {Pakmor}, {Weinberger}, {Vogelsberger}, {Marinacci}, \&
  {Hernquist}}]{nelson20}
{Nelson}, D., {Sharma}, P., {Pillepich}, A., {et~al.} 2020, \mnras, 498, 2391

\bibitem[{{Olivares} {et~al.}(2019){Olivares}, {Salome}, {Combes}, {Hamer},
  {Guillard}, {Lehnert}, {Polles}, {Beckmann}, {Dubois}, {Donahue}, {Edge},
  {Fabian}, {McNamara}, {Rose}, {Russell}, {Tremblay}, {Vantyghem}, {Canning},
  {Ferland}, {Godard}, {Peirani}, \& {Pineau des Forets}}]{olivares19}
{Olivares}, V., {Salome}, P., {Combes}, F., {et~al.} 2019, \aap, 631, A22

\bibitem[{{Omma} {et~al.}(2004){Omma}, {Binney}, {Bryan}, \& {Slyz}}]{omma04}
{Omma}, H., {Binney}, J., {Bryan}, G., \& {Slyz}, A. 2004, \mnras, 348, 1105

\bibitem[{{Oosterloo} {et~al.}(2023){Oosterloo}, {Morganti}, \&
  {Murthy}}]{oosterloo23}
{Oosterloo}, T., {Morganti}, R., \& {Murthy}, S. 2023, arXiv e-prints,
  arXiv:2312.00917

\bibitem[{{Perucho} {et~al.}(2019){Perucho}, {Mart{\'\i}}, \&
  {Quilis}}]{perucho19}
{Perucho}, M., {Mart{\'\i}}, J.-M., \& {Quilis}, V. 2019, \mnras, 482, 3718

\bibitem[{{Pope} {et~al.}(2010){Pope}, {Babul}, {Pavlovski}, {Bower}, \&
  {Dotter}}]{pope10}
{Pope}, E.~C.~D., {Babul}, A., {Pavlovski}, G., {Bower}, R.~G., \& {Dotter}, A.
  2010, \mnras, 406, 2023

\bibitem[{{Qiu} {et~al.}(2020){Qiu}, {Bogdanovi{\'c}}, {Li}, {McDonald}, \&
  {McNamara}}]{qiu20}
{Qiu}, Y., {Bogdanovi{\'c}}, T., {Li}, Y., {McDonald}, M., \& {McNamara}, B.~R.
  2020, Nature Astronomy, 4, 900

\bibitem[{{Qiu} {et~al.}(2019){Qiu}, {Bogdanovi{\'c}}, {Li}, {Park}, \&
  {Wise}}]{qiu19}
{Qiu}, Y., {Bogdanovi{\'c}}, T., {Li}, Y., {Park}, K., \& {Wise}, J.~H. 2019,
  \apj, 877, 47

\bibitem[{{Revaz} {et~al.}(2008){Revaz}, {Combes}, \& {Salom{\'e}}}]{revaz08}
{Revaz}, Y., {Combes}, F., \& {Salom{\'e}}, P. 2008, \aap, 477, L33

\bibitem[{{Reynolds} {et~al.}(2005){Reynolds}, {McKernan}, {Fabian}, {Stone},
  \& {Vernaleo}}]{reynolds05}
{Reynolds}, C.~S., {McKernan}, B., {Fabian}, A.~C., {Stone}, J.~M., \&
  {Vernaleo}, J.~C. 2005, \mnras, 357, 242

\bibitem[{{Russell} {et~al.}(2017){Russell}, {McNamara}, {Fabian}, {Nulsen},
  {Combes}, {Edge}, {Hogan}, {McDonald}, {Salom{\'e}}, {Tremblay}, \&
  {Vantyghem}}]{russell17}
{Russell}, H.~R., {McNamara}, B.~R., {Fabian}, A.~C., {et~al.} 2017, \mnras,
  472, 4024

\bibitem[{{Russell} {et~al.}(2019){Russell}, {McNamara}, {Fabian}, {Nulsen},
  {Combes}, {Edge}, {Madar}, {Olivares}, {Salom{\'e}}, \&
  {Vantyghem}}]{russell19}
---. 2019, \mnras, 490, 3025

\bibitem[{{Saikia}(2022)}]{saikia22}
{Saikia}, D.~J. 2022, Journal of Astrophysics and Astronomy, 43, 97

\bibitem[{{Salom{\'e}} {et~al.}(2008){Salom{\'e}}, {Revaz}, {Combes}, {Pety},
  {Downes}, {Edge}, \& {Fabian}}]{salom08}
{Salom{\'e}}, P., {Revaz}, Y., {Combes}, F., {et~al.} 2008, \aap, 483, 793

\bibitem[{{Salom{\'e}} {et~al.}(2006){Salom{\'e}}, {Combes}, {Edge},
  {Crawford}, {Erlund}, {Fabian}, {Hatch}, {Johnstone}, {Sanders}, \&
  {Wilman}}]{salom06}
{Salom{\'e}}, P., {Combes}, F., {Edge}, A.~C., {et~al.} 2006, \aap, 454, 437

\bibitem[{{Saxton} {et~al.}(2001){Saxton}, {Sutherland}, \&
  {Bicknell}}]{saxton01}
{Saxton}, C.~J., {Sutherland}, R.~S., \& {Bicknell}, G.~V. 2001, \apj, 563, 103

\bibitem[{{Schure} {et~al.}(2009){Schure}, {Kosenko}, {Kaastra}, {Keppens}, \&
  {Vink}}]{schure09}
{Schure}, K.~M., {Kosenko}, D., {Kaastra}, J.~S., {Keppens}, R., \& {Vink}, J.
  2009, \aap, 508, 751

\bibitem[{{Simionescu} {et~al.}(2009){Simionescu}, {Werner}, {B{\"o}hringer},
  {Kaastra}, {Finoguenov}, {Br{\"u}ggen}, \& {Nulsen}}]{simionescu09}
{Simionescu}, A., {Werner}, N., {B{\"o}hringer}, H., {et~al.} 2009, \aap, 493,
  409

\bibitem[{{Su} {et~al.}(2021){Su}, {Hopkins}, {Bryan}, {Somerville}, {Hayward},
  {Angl{\'e}s-Alc{\'a}zar}, {Faucher-Gigu{\`e}re}, {Wellons}, {Stern},
  {Terrazas}, {Chan}, {Orr}, {Hummels}, {Feldmann}, \&
  {Kere{\v{s}}}}]{sukungyi21}
{Su}, K.-Y., {Hopkins}, P.~F., {Bryan}, G.~L., {et~al.} 2021, \mnras, 507, 175

\bibitem[{{Sutherland} \& {Dopita}(1993)}]{sd93}
{Sutherland}, R.~S., \& {Dopita}, M.~A. 1993, \apjs, 88, 253

\bibitem[{{Tamhane} {et~al.}(2022){Tamhane}, {McNamara}, {Russell}, {Edge},
  {Fabian}, {Nulsen}, \& {Babyk}}]{tamhane22}
{Tamhane}, P.~D., {McNamara}, B.~R., {Russell}, H.~R., {et~al.} 2022, \mnras,
  516, 861

\bibitem[{{Tamhane} {et~al.}(2023){Tamhane}, {McNamara}, {Russell}, {Combes},
  {Qiu}, {Edge}, {Maiolino}, {Fabian}, {Nulsen}, {Johnstone}, \&
  {Carniani}}]{tamhane23}
---. 2023, \mnras, 519, 3338

\bibitem[{{Townsend}(2009)}]{townsend09}
{Townsend}, R.~H.~D. 2009, \apjs, 181, 391

\bibitem[{{Valentini} \& {Brighenti}(2015)}]{valentini15}
{Valentini}, M., \& {Brighenti}, F. 2015, \mnras, 448, 1979

\bibitem[{{Vernaleo} \& {Reynolds}(2006)}]{vernaleo06}
{Vernaleo}, J.~C., \& {Reynolds}, C.~S. 2006, \apj, 645, 83

\bibitem[{{Weinberger} {et~al.}(2018){Weinberger}, {Springel}, {Pakmor},
  {Nelson}, {Genel}, {Pillepich}, {Vogelsberger}, {Marinacci}, {Naiman},
  {Torrey}, \& {Hernquist}}]{weinberger18}
{Weinberger}, R., {Springel}, V., {Pakmor}, R., {et~al.} 2018, \mnras, 479,
  4056

\bibitem[{{Weinberger} {et~al.}(2023){Weinberger}, {Su}, {Ehlert}, {Pfrommer},
  {Hernquist}, {Bryan}, {Springel}, {Li}, {Burkhart}, {Choi}, \&
  {Faucher-Gigu{\`e}re}}]{weinberger23}
{Weinberger}, R., {Su}, K.-Y., {Ehlert}, K., {et~al.} 2023, \mnras, 523, 1104

\bibitem[{{Xia} {et~al.}(2018){Xia}, {Teunissen}, {El Mellah}, {Chan{\'e}}, \&
  {Keppens}}]{xia18}
{Xia}, C., {Teunissen}, J., {El Mellah}, I., {Chan{\'e}}, E., \& {Keppens}, R.
  2018, \apjs, 234, 30

\bibitem[{{Zhang} {et~al.}(2022){Zhang}, {Zhuravleva}, {Gendron-Marsolais},
  {Churazov}, {Schekochihin}, \& {Forman}}]{zhang22}
{Zhang}, C., {Zhuravleva}, I., {Gendron-Marsolais}, M.-L., {et~al.} 2022,
  \mnras, 517, 616

\end{thebibliography}
\end{document}